\documentclass{aa}

\usepackage[normalem]{ulem}
\usepackage[dvipsnames]{xcolor}
\usepackage{xspace}
\usepackage{amsmath}
\usepackage{rotating}
\usepackage{threeparttable}
\usepackage{enumerate}
\usepackage{color}
\usepackage{graphicx}
\usepackage{soul}
\usepackage{txfonts}
\usepackage{hyperref}
\usepackage{stfloats}
\usepackage{orcidlink}

\newcommand{\hstfull}{\textit{Hubble Space Telescope}\xspace}
\newcommand{\hst}{\textit{HST}\xspace}
\newcommand{\jwst}{\textit{JWST}\xspace}
\newcommand{\gaia}{\textit{Gaia}\xspace}

\newcommand{\maspixel}{mas pixel$^{-1}$\xspace}

\defcitealias{2025A&A...698A.197M}{Paper~I}
\defcitealias{2026rosignoli}{Paper~II}

\begin{document} 

\title{The \textit{Hubble} Missing Globular Cluster Survey}
\subtitle{III. Astro-photometric catalogs, artificial-star tests, and improved absolute proper motions}

\titlerunning{The \hstfull Missing Globular Cluster Survey. III.}
\authorrunning{Libralato et al.}

\author{M. Libralato\inst{1}\orcidlink{0000-0001-9673-7397}
    \and
    A. Bellini\inst{2}\orcidlink{0000-0003-3858-637X}
    \and
    D. Massari\inst{3}\orcidlink{0000-0001-8892-4301}
    \and
    M. Bellazzini\inst{3}\orcidlink{0000-0001-8200-810X}
    \and
    F. Aguado-Agelet\inst{4,5}\orcidlink{0000-0003-2713-1943}
    \and
    S. Cassisi\inst{6,7}\orcidlink{0000-0001-5870-3735}
    \and
    E. Ceccarelli\inst{3,8}\orcidlink{009-0007-3793-9766}
    \and
    E. Dalessandro\inst{3}\orcidlink{0000-0003-4237-4601}
    \and 
    E. Dodd\inst{9}\orcidlink{0000-0002-2691-7728}
    \and 
    F. R. Ferraro\inst{8}\orcidlink{0000-0002-2165-8528}
    \and
    C. Gallart\inst{5,10}\orcidlink{0000-0001-6728-806X}
    \and 
    B. Lanzoni\inst{8}\orcidlink{0000-0001-5613-4938}
    \and
    M. Monelli\inst{6,5,10}\orcidlink{0000-0001-5292-6380}
    \and
    A. Mucciarelli\inst{8,3}\orcidlink{0000-0001-9158-8580}
    \and
    E. Pancino\inst{11}\orcidlink{0000-0003-0788-5879}
    \and
    R. Pascale\inst{3}\orcidlink{0000-0002-6389-6268}
    \and
    L. Rosignoli\inst{8,3}\orcidlink{0000-0002-0327-5929}
    \and
    M. Salaris\inst{12,6}\orcidlink{0000-0002-2744-1928}
    \and
    S. Saracino\inst{11,12}\orcidlink{0000-0003-4746-6003}
    \and
    C. Zerbinati\inst{3,8}\orcidlink{0009-0004-2797-4056}
    }

\institute{
INAF - Osservatorio Astronomico di Padova, Vicolo dell'Osservatorio 5, Padova I-35122, Italy \\
\email{mattia.libralato@inaf.it}
\and
Space Telescope Science Institute, 3700 San Martin Drive, Baltimore, MD 21218, USA
\and
INAF - Astrophysics and Space Science Observatory of Bologna, Via Gobetti 93/3, 40129 Bologna, Italy
\and
atlanTTic, Universidade de Vigo, Escola de Enxeñar\'ia de Telecomunicaci\'on, 36310, Vigo, Spain 
\and
Universidad de La Laguna, Avda. Astrof\'isico Fco. S\'anchez, E-38205 La Laguna, Tenerife, Spain
\and
INAF – Osservatorio Astronomico di Abruzzo, Via M. Maggini, 64100 Teramo, Italy
\and
INFN - Sezione di Pisa, Universit\'a di Pisa, Largo Pontecorvo 3, 56127 Pisa, Italy
\and
Department of Physics and Astronomy, University of Bologna, Via Gobetti 93/2, 40129 Bologna, Italy      
\and
Institute for Computational Cosmology \& Centre for Extragalactic Astronomy, Department of Physics, Durham University, South Road, Durham, DH1 3LE, UK
\and
Instituto de Astrofísica de Canarias, Calle V\'ia L\'actea s/n, 38206 La Laguna, Tenerife, Spain
\and
INAF - Osservatorio Astrofisico di Arcetri, Largo E. Fermi 5, I-50125 Firenze, Italy
\and
Astrophysics Research Institute, Liverpool John Moores University, 146 Brownlow Hill, Liverpool L3 5RF, UK
}

\date{Received 07 February 2026 / Accepted 21 March 2026}
 
\abstract
{
The \textit{Hubble} Missing Globular Cluster Survey (MGCS) has taken one of the last opportunities to complete the census of Galactic globular clusters (GCs) started by past \hstfull (\hst) programs, securing high-resolution data for 34 GCs never observed before by \hst. The previous papers in the series have highlighted the astrometric and photometric potential of the project by analyzing a subsample of targets. We present, and release to the community, the official astro-photometric catalogs of the MGCS for all GCs imaged by this project. We describe the data reduction using state-of-the-art techniques designed for \hst. We discuss the photometric calibration and show, for the first time, the synergy with the \gaia catalog to ensure homogeneous photometry across our data set. We compute artificial-star tests that can be used to assess systematics and the completeness level of our data. We combined \hst and \gaia data to refine the absolute proper motions of our GCs, reaching a precision $\sim$3 times better than that of \gaia alone. We used these new proper motions to update (and to determine for the first time for six systems) the associations between GCs and their putative galaxy progenitors. This work continues decades-long efforts of large Treasury programs in sharing precise and accurate atlases to the community for studying GCs across a wide range of scientific endeavors.
}


\maketitle

\nolinenumbers

\section{Introduction}

At the heart of any advancement in astrophysics lies the crucial need to answer two fundamental questions: ``where'' we test our theories and ``how'' we improve our understanding of a phenomenon. For stellar populations, globular clusters (GCs) constitute one of the best choices to fill the role of the where, providing ideal laboratories. The how is equally critical, and sometimes more challenging to establish. For decades, the stability, resolution, and depth of the \hstfull (\hst) have made this telescope the best tool for this field. Over time, \hst has become instrumental not only for the detailed analysis of individual GCs, but also for the study of their characteristics as a system of Galactic tracers. To enable these broad and impactful \hst-based studies on the topic, a homogeneous framework that encompasses consistent photometric catalogs and theoretical models was essential. Extensive precious Treasury and Legacy Survey programs with \hst have already demonstrated that homogeneity streamlines and simplifies analyzes and comparisons of clusters. For example, almost 20 years ago, the ``ACS Survey of Galactic Globular Clusters'' \citep[GO-10775][]{2007AJ....133.1658S} imaged the cores of 65 GCs (and one open cluster) to construct a uniform database of deep F606W and F814W photometry \citep{2008AJ....135.2055A}. Later, when the presence of the multiple stellar populations became irrefutable, the ``\hst UV Legacy Survey of Galactic Globular Clusters'' \citep[GO-13297]{2015AJ....149...91P} was designed to photometrically shed light on the phenomenon. Similarly to the GO-10775 survey, the GO-13927 program ultraviolet and optical campaign released homogeneous photometric \citep{2018MNRAS.481.3382N} and astrometric \citep{2022LibralatoPMcat} catalogs. The combination of just these two \hst programs has led to hundreds of publications and a leap forward in the understanding of GCs. The \textit{Hubble} Missing Globular Cluster Survey (MGCS) is meant to continue this legacy, adding new clusters to the sample.

The MGCS is an \hst Treasury program (GO-17435; PI: Massari) dedicated to observing kinematically-confirmed Galactic globular clusters that lacked suitable \hst imaging data \citep[hereafter Paper~I]{2025A&A...698A.197M}. The project aims at increasing our overall knowledge of GCs and the Milky Way (MW) in general, and its key objectives include accurate age determinations of the GC \citep[that, combined with astrometric and spectroscopic information, are essential to understand GC origin and the main MW formation events; e.g.,][]{2019A&A...630L...4M,kruijssen20, 2020ARA&A..58..205H,2021MNRAS.500.1385H,2022A&A...663A..38K, callingham22, 2023A&A...680A..20M,chen24, ceccarelli25, niederhofer25}, the search for MW bulge relics, and the inference of structural and dynamical properties of GCs.  In this work, we present the astro-photometric catalogs of the MGCS and make them publicly available to the community. We summarize the data reduction and the various steps taken to ensure photometric homogeneity in Sect.~\ref{sec:datared}. We then describe the artificial-star tests included in the release (Sect.~\ref{sec:art}). Finally, we present an application made possible by our data set (Sect.~\ref{sec:pm}). Specifically, we link our astro-photometric catalogs to the \gaia Data Release 3 (DR3) catalog \citep{2016GaiaCit,2023A&A...674A...1G} and compute proper motions (PMs) for the stars in common, refining and improving (in terms of precision) the absolute PMs for most clusters in our sample.\looseness=-4

\section{Data reduction}\label{sec:datared}

Our sample comprises 34 clusters, 27 of which were observed through the Wide Field Channel (WFC) of the Advanced Camera for Surveys (ACS) and the Ultraviolet and Visible channel (UVIS) of Wide-Field Camera 3 (WFC3), and the remaining seven systems were imaged with the Infra-Red (IR) channel of the WFC3\footnote{The data sets used in this work are collected at \href{https://doi.org/10.17909/ncwh-db07}{https://doi.org/10.17909/ncwh-db07}\,.}. A complete log of observations is detailed in \citetalias{2025A&A...698A.197M}. As described in \citet{2025A&A...698A.197M}, the GC sample of the MGCS was selected as a compromise between technical requirements (moderate distance and reddening) and scientific requirements (systems that occupy regions of the parameter space--mass, density, environment, which are rather extreme and poorly sampled).

\begin{figure*}
    \centering
    \includegraphics[width=\textwidth]{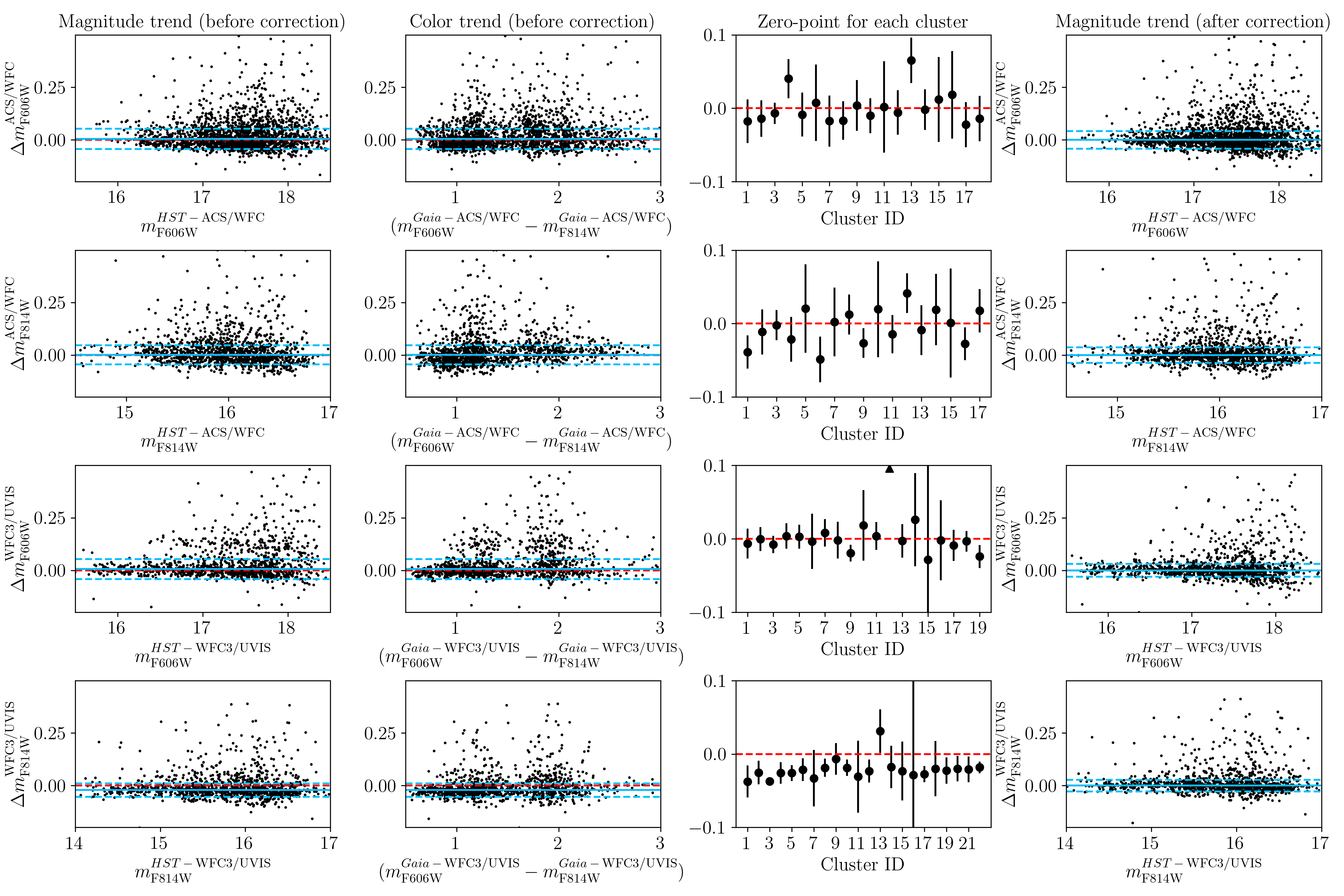}
    \caption{Comparison of our photometry with that in the \gaia synthetic-photometry catalog. In each row ($=$camera and filter), the first two panels from the left show the \hst-\gaia magnitude difference for all stars in all our GCs as a function of magnitude and color, respectively. The red line is set to 0 as a reference, while the light-blue lines are set at the median (solid line) $\pm$ 1$\sigma$ (dashed lines) magnitude difference for all GCs. The third panel from the left shows the average zero-point difference for each cluster. Finally, the last panel on the right shows the magnitude trend after the photometric correction.}
    \label{fig:zp}
\end{figure*}

The ACS+WFC3 campaign was originally designed to collect data in two back-to-back visits to provide images in two filters of both the core (with ACS/WFC) and the outskirts (with WFC3/UVIS) of each target. However,  sporadic visit failures\footnote{Mostly due to guide-star acquisition issues.} compromised this setup for some targets, and we requested a repeat of the failed visits through the \hst Operations Problem Report (HOPR). In order to improve the chances for the repeats to be successful, we had to change the originally planned telescope orientation that, in turn, resulted in a partial or null overlap of the parallel fields between the two visits. This is the reason why some clusters do not have two-filter photometry in the WFC3/UVIS parallel fields.

The astro-photometric catalogs were obtained following the prescriptions extensively described in many investigations on the same topic with \hst data \citep[e.g.,][and references therein]{2017BelliniwCenI,2018LibralatoNGC362,2018MNRAS.481.3382N,2022LibralatoPMcat,2024A&A...689A.162N}, and consist of multiple steps. In the following, we briefly describe the main steps.

First, we made use of calibrated\footnote{Calibrated exposures are dark- and/or bias-subtracted, flat-fielded, gain-corrected and, for the optical sample, also corrected for charge-transfer-efficiency defects as detailed in \citet{2018acs..rept....4A}.}, unresampled images, which are best suited for astro-photometric analyzes based on point spread function (PSF) fitting techniques like the one we performed in our project. For each image, we obtained a preliminary list of the positions and fluxes of bright and isolated sources by fitting a set of effective-PSF (ePSF) models \citep[first-pass photometry;][]{2022acs..rept....2A}. These ePSFs were tailored to each exposure starting from a set of library ePSF models\footnote{\href{https://www.stsci.edu/~jayander/HST1PASS/LIB/PSFs/}{https://www.stsci.edu/$\sim$jayander/HST1PASS/LIB/PSFs/}} to account for variations due to telescope breathing \citep[see, e.g., Section 3 of ][]{2017BelliniwCenI}. The positions were also corrected for the effects of geometric distortion by means of the available\footnote{\href{https://www.stsci.edu/~jayander/HST1PASS/LIB/GDCs/}{https://www.stsci.edu/$\sim$jayander/HST1PASS/LIB/GDCs/}} corrections for the ACS/WFC \citep{2006acs..rept....1A}, WFC3/UVIS \citep{2009BelliniWFC3,2011BelliniWFC3} and WFC3/IR \citep{2016AndersonWFC3IRPSF} cameras.

Second, we built a common reference frame in which all images could be properly transformed and analyzed simultaneously \citep[see, e.g.,][for more details]{2017BelliniwCenI,2018LibralatoNGC362}. For astrometric registration, we started from the \gaia Early DR3 (EDR3) catalog \citet{2016GaiaCit,2021GaiaEDR3}. For each cluster, we projected the \gaia catalog onto a tangent plane centered on the center of the cluster taken from the literature \citepalias[see][]{2025A&A...698A.197M}. The positions of the sources in the \gaia catalog were propagated using the available PMs to the average epoch of our \hst observations. This step allowed us to anchor our reference frame onto the absolute frame of the International Celestial Reference System (ICRS). Although the axis orientation ($x$ pointing to the west and $y$ to the north, respectively) and scale (40 \maspixel and 130 \maspixel for the GCs in the optical and IR samples, respectively) were kept consistent across all clusters in the same sample (optical or IR), the center of the cluster in the pixel-based frame was shifted to ensure non-negative coordinates. From the photometric point of view, instrumental magnitudes\footnote{$m_{\rm instrumental} = -2.5\log_{10}({\rm Flux})$} were zero-pointed to the longest exposure in each camera and filter.

Third, we created the so-called bright-star list. This preliminary catalog includes only very bright sources, both saturated and not, and serves a twofold purpose: (i) it marks the position of bright objects with pronounced diffuse light, diffraction spikes, and sometimes even bleeding columns (in optical detectors), which need to be masked off in the second-pass photometry stage; and (ii) it keeps a record (position and magnitude) of saturated sources that are not measured in the second-pass photometry, so that they could still be included in the final catalog.

As a fourth and last step, we ran the second-pass photometry with the code \texttt{KS2} \citep{2017BelliniwCenI,2018MNRAS.481.3382N,2018LibralatoNGC362,2022LibralatoPMcat}, which is designed to measure neighbor-subtracted stars in crowded environments like GCs, and to enhance the detectability of faint sources by using all exposures simultaneously during the detection phase. The \texttt{KS2} code measures and subtracts progressively fainter objects in various stages, the detection criteria of which can be fine-tuned for any given set of observations \citep{2017BelliniwCenI,2022LibralatoPMcat}. As exhaustively described in the aforementioned literature, there is no one-size-fits-all recipe to optimally run \texttt{KS2}. For our work, we found a set of parameters that provided a reasonable compromise between the completeness of the catalog and the reliability of the detected sources, and kept the same parameters for all clusters in our sample \citep[for a detailed description of the various setup parameters, we refer interested readers to][]{2017BelliniwCenI}.

The final catalog from \texttt{KS2} contains positions and magnitudes of detected sources, as well as a series of quality parameters that can be used to select a bona fide sample of well-measured sources. Saturated stars from the first-pass photometry and present in the bright list not measured by \texttt{KS2} are also included in the final catalog\footnote{Note that the bright-star list might occasionally include unsaturated stars that were not measured by the second-pass-photometry tool. These sources have been included in the final catalog and flagged as saturated to alert the user, as these objects are likely spurious detections that we only included for completeness.}.

As anticipated in \citetalias{2025A&A...698A.197M}, the pixel-based positions were also transformed into ICRS Equatorial coordinates by means of the \gaia EDR3 catalog. Our photometry was calibrated in the Vega-magnitude flight system as follows:\ (i) we ran aperture photometry (with aperture radius selected on a case-by-case basis) on the drizzled images (\texttt{\_drc} for ACS/WFC and WFC3/UVIS, and \texttt{\_drz} for WFC3/IR data, respectively); (ii) we corrected the drizzled-based photometry to account for the finite aperture by using the official aperture corrections provided for the instruments;\footnote{See the official aperture-correction pages: \href{https://www.stsci.edu/hst/instrumentation/acs/data-analysis/aperture-corrections}{ACS/WFC}, \href{https://www.stsci.edu/hst/instrumentation/wfc3/data-analysis/photometric-calibration/}{WFC3}.}(iii) we cross-identified bright, unsaturated stars in common between our \texttt{KS2} and drizzled-based catalogs, and computed the 2.5$\sigma$-clipped average magnitude between the two photometric sets ($\Delta m$); (iv) the final calibrated magnitudes are defined as: $m_{\rm Vega-mag} = m_{\rm instrumental} + \Delta m + {\rm ZP_{Vega-mag}}$, where ${\rm ZP_{Vega-mag}}$ are the camera-dependent official Vega-mag zero-points.\footnote{See the official photometric-calibration pages: \href{https://www.stsci.edu/hst/instrumentation/acs/data-analysis/zeropoints}{ACS/WFC}, \href{https://www.stsci.edu/hst/instrumentation/wfc3/data-analysis/photometric-calibration}{WFC3}.}

A visual inspection of the color-magnitude diagrams (CMDs) based on ACS/WFC and WFC3/UVIS data revealed the presence of photometric discrepancies between the two sets in some clusters. While differences on the order of a few hundredths of a magnitude (i.e., of the order of the calibration errors) are expected, we found cases where the photometric offset was as high as 0.1 mag. Due to the environment of some systems (e.g., patchy or high-extinction regions), it was not trivial to identify the culprit in such discrepancies (whether they were systematic errors in the calibration or genuine reddening features) and determine which photometric set (ACS, WFC3, or both) was affected by this systematic. For this reason, we performed an external cross-check on our photometry by comparing it with the synthetic photometry of the \gaia DR3 low-resolution spectra \citep{2023A&A...674A...2D} retrieved with the \texttt{GaiaXPy}\footnote{\href{https://gaia-dpci.github.io/GaiaXPy-website/}{https://gaia-dpci.github.io/GaiaXPy-website/}} tool.

For the GCs in the optical sample, we cross-identified the same stars in our \texttt{KS2} and in the \gaia synthetic-photometry catalogs. We then computed the difference between our \hst-based and the \gaia synthetic photometry. The result is presented in Fig.~\ref{fig:zp}. We find an overall agreement between the two sets for all filters but the WFC3/UVIS F814W, where the average zero-point between the \hst and \gaia catalogs is about $-0.02$ mag. This offset is on the same order as the typical uncertainty in the VEGAmag calibration \citep[0.01--0.03 mag; see, e.g.,][]{2018MNRAS.481.3382N}. We cannot rule out \gaia synthetic photometry as the source of this discrepancy, but this is unlikely given that we do not observe similar zero-point offsets in other filters derived from the same spectra. We also do not find a clear magnitude or color trend. To homogenize the photometry for all GCs as much as possible, for each cluster we applied to our calibrated optical photometry the corresponding zero-point coming from the \gaia comparison. For 11 GCs with less than five stars in common between the \hst catalog (in at least one filter) and the \gaia synthetic-photometry catalog, we applied the average zero-point coming from the entire sample of clusters. Table~\ref{tab:phot} provides the median values (and standard errors) of the differences between our \hst and \gaia photometry before and after correction. Note that this correction does not remove all spurious systematic offsets in our photometry, as \gaia sometimes struggles in fields with high reddening and/or crowding or because of the small statistics, but it is another step towards obtaining a photometrically-homogeneous sample of GCs.

The astro-photometric catalogs obtained with second-pass photometry, which contain positions, magnitudes, and quality parameters (that can be used to select the best stars for analysis and remove spurious detections) are described in Appendix~\ref{appendix:release}. We also make publicly available the astrometrized stacked images (one for each filter, camera, exposure time).

\begin{table}[]
    \centering
    \caption{Comparison between our \hst and the \gaia synthetic-photometry catalog.}\label{tab:phot}
    \begin{tabular}{r|c|c}
        \hline
        \hline
            Camera \& Filter & $\Delta m$ (Before) & $\Delta m$ (After) \\
        \hline
        \hline
           ACS/WFC F606W & $ 0.004 \pm 0.001$ & $0.000 \pm 0.001$ \\
           ACS/WFC F814W & $ 0.001 \pm 0.001$ & $0.000 \pm 0.001$ \\
         WFC3/UVIS F606W & $ 0.006 \pm 0.001$ & $0.000 \pm 0.001$ \\
         WFC3/UVIS F814W & $-0.023 \pm 0.001\phantom{-}$ & $0.000 \pm 0.001$ \\
        \hline
        \hline
    \end{tabular}
    \tablefoot{We provide the median and standard error of the magnitude difference between our \hst and the \gaia synthetic photometry before and after the photometric correction described in the text.}
\end{table}

\begin{figure*}
    \centering
    \includegraphics[width=\textwidth]{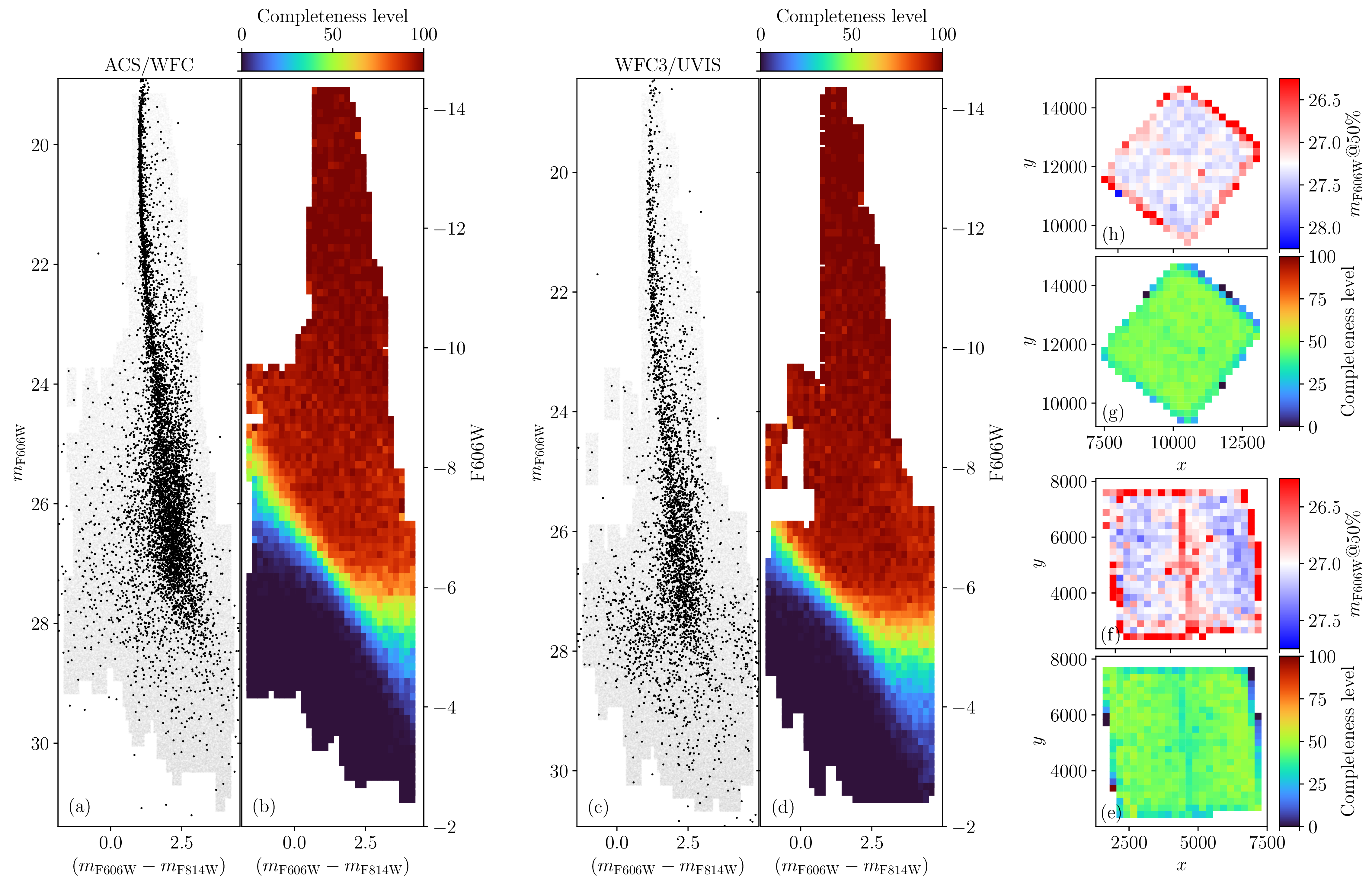}
    \caption{Overview of the artificial-star tests for the GC ESO-452-11. Panel (a) shows the input CMD for the ACS/WFC data. Black dots are taken from the observed catalog, while gray points are artificial stars in input. Panel (b) presents the Hess diagram of recovered stars, color-coded by the completeness level as in the colorbar at the top. The analogs of these plots for the WFC3/UVIS parallel field are provided in panels (c) and (d). Panels (e) and (f) display the Hess diagrams of the spatial completeness in the ACS/WFC region. The former is color-coded according to the average completeness level, while the latter is color-coded according to the F606W magnitude at which the completeness level reaches the 50\%. The $x$- and $y$-axis units are WFC3/UVIS pixels (units in the plot labels are omitted for clarity). Similarly, panels (g) and (h) show the spatial completeness in the WFC3/UVIS region.}
    \label{fig:artstars2D}
\end{figure*}

\section{Artificial-star tests}\label{sec:art}

We performed homogeneous artificial-star injection and retrieval on all available data sets. The resulting ancillary catalogs can be used to assess the level of completeness in our data and to search and correct for potential systematic errors \citep[as in, e.g.,][]{2009ApJ...697..965B,2024A&A...690A.371L}. The observed CMDs of the GCs in our program revealed features and sequences related to the various populations present in the images (cluster and field stars). As such, we generated lists of artificial stars that span a wide range of colors and magnitudes covering the color-magnitude space of real sources \citep{2017BelliniwCenI,2023NardiellojwstIII,2023LibralatoNIRISS} instead of adopting a single fiducial line \citep{2008AJ....135.2055A}.

For each GC in the ACS$+$WFC3 sample, we started by constructing a Hess diagram of the observed CMD. We considered all stars between instrumental magnitude $-14.25$ ($\sim$0.5 mag above the saturation threshold of the long exposures, to also test the completeness in the magnitude regime only covered by short exposures) and $-2.5$ (at about the level of the faintest sources detected by \texttt{KS2}). When a cell (with size $\sim$0.2 in magnitude and $\sim$0.1 in color) of the Hess diagram contained at least 5 stars, we generated a set of artificial stars with random magnitude and color within the cell boundaries. The number of artificial stars per cell quadratically increases between 50 (bright-end of the CMD) and 250 (faint-end of the CMD), to better account for incompleteness effects in the faint regime. In the central ACS/WFC field, the associated random positions were selected from either flat (like that of field objects) or Gaussian distributions (centered in the cluster's center and with $\sigma$ of 1500 pixels) with a 50-50 ratio; only a flat distribution was considered in the parallel, field-dominated WFC3/UVIS regions. In addition, the positions of the stars were chosen so that they landed in at least one \hst exposure and, for the WFC3/UVIS fields, in a region covered by two filters. Due to this empirical approach, the number of entries in the artificial-star catalogs is not exactly the same for all GCs and can range from a minimum of $\sim$93\,000 (Whiting~1) to a maximum of $\sim$393\,000 (ESO-280-6). A similar approach was also used for the artificial stars of GCs in the WFC3/IR sample (simulated between $\sim$243\,000 and $\sim$292\,000 stars).

These artificial star lists were used as input in new runs of \texttt{KS2}. Each star was added to each image one at a time, measured in the same way as the real stars, and then removed, to not bias the measurement of other artificial stars. In the following, we considered a star as recovered if the input and output positions differ by less than 0.5 pixel, and magnitudes agree to within 0.75 mag (i.e., fluxes agree within a factor of 2). Our artificial-star catalogs list input and measured positions and magnitudes, so users are free to apply different criteria that best suit their needs.

\begin{figure}
    \centering
    \includegraphics[width=\columnwidth]{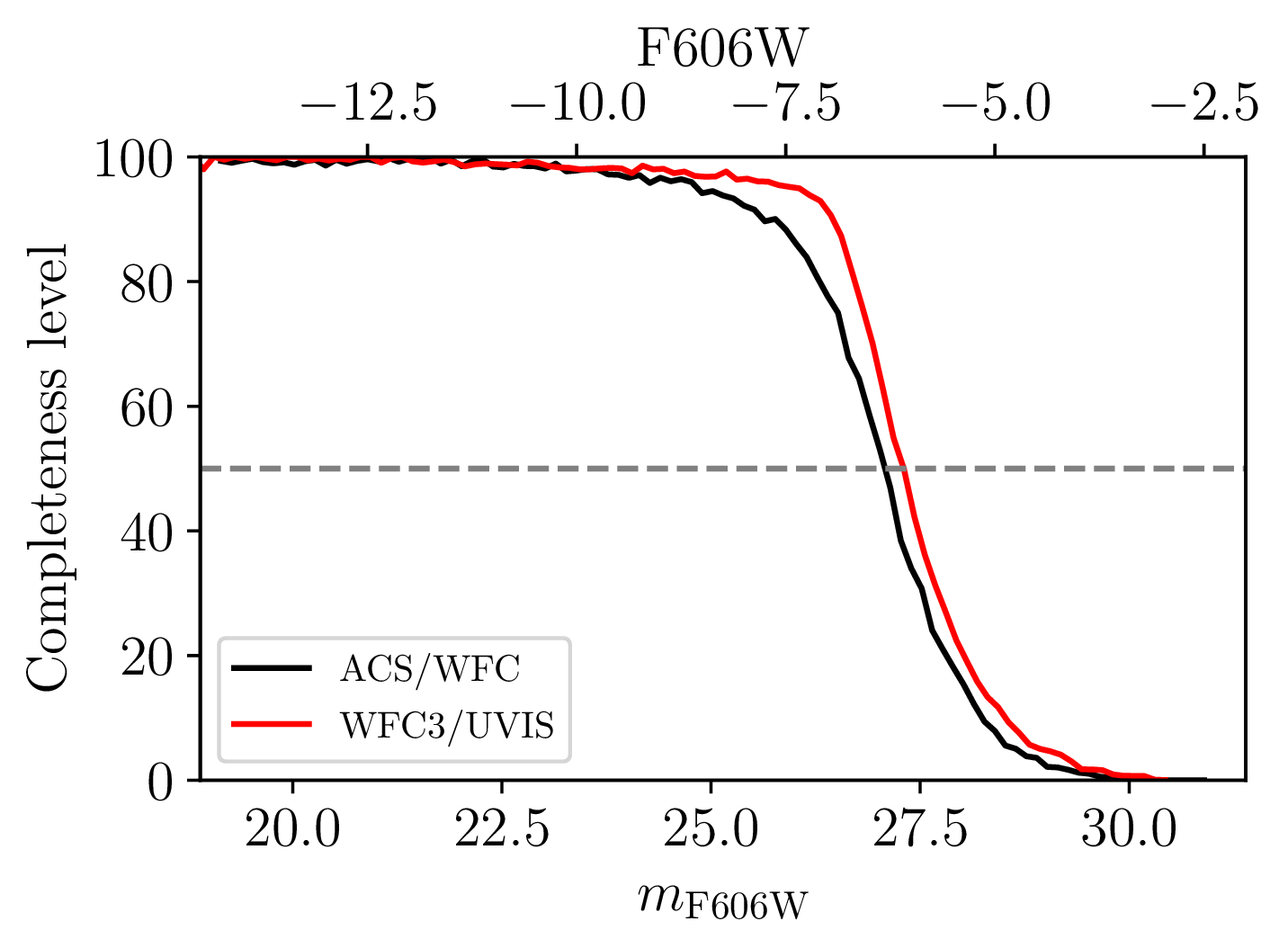}
    \caption{Completeness level for the GC ESO-452-11 as a function of F606W magnitude (instrumental at the top and calibrated at the bottom) for the ACS/WFC (black line) and WFC3/UVIS (red line) data. The 50\% completeness level is highlighted with a gray, dashed line.}
    \label{fig:artstars}
\end{figure}

Figure~\ref{fig:artstars2D} illustrates the inputs, outputs, and statistics from the artificial-star test of the GC ESO-452-11. Panels (a) and (c) present the input CMDs for the ACS/WFC and WFC3/UVIS fields, respectively. The black dots are the sources from the observed catalog, while gray points represent the artificial stars in input. Panels (b) and (d) show the Hess diagrams of the recovered CMD, color-coded by the completeness level as indicated in the corresponding colorbar at the top. The 50\% completeness level is reached, on average, at $m_{\rm F606W} \sim 27.5$, i.e., 7 magnitudes below the main-sequence turn-off. The Hess diagrams in panels (e) and (g) depict the average spatial completeness of the entire catalog (again, color-coded as in the side colorbars). Panels (f) and (g) display the Hess diagrams in which cells are color-coded according to the F606W magnitude at which the completeness reaches the 50\% level. For this GC, the completeness is spatially uniform, with a slightly lower level towards the edges of the field, due to the dithering strategy of the observations. The completeness level also progressively drops in the proximity of the chip gap in the ACS/WFC field as a result of higher uncorrected residuals related to the charge-transfer inefficiency of the detector \citep{Kuhn2024}. Finally, Fig.~\ref{fig:artstars} presents a more classic view of the completeness of the F606W data in the ACS/WFC (black line) and WFC3/UVIS fields (red line).

We made our artificial-star tests (both input and output catalogs) available to the community. A description of these catalogs is provided in Appendix~\ref{appendix:release}.

\begin{figure*}
    \centering
    \includegraphics[width=9.2cm, keepaspectratio]{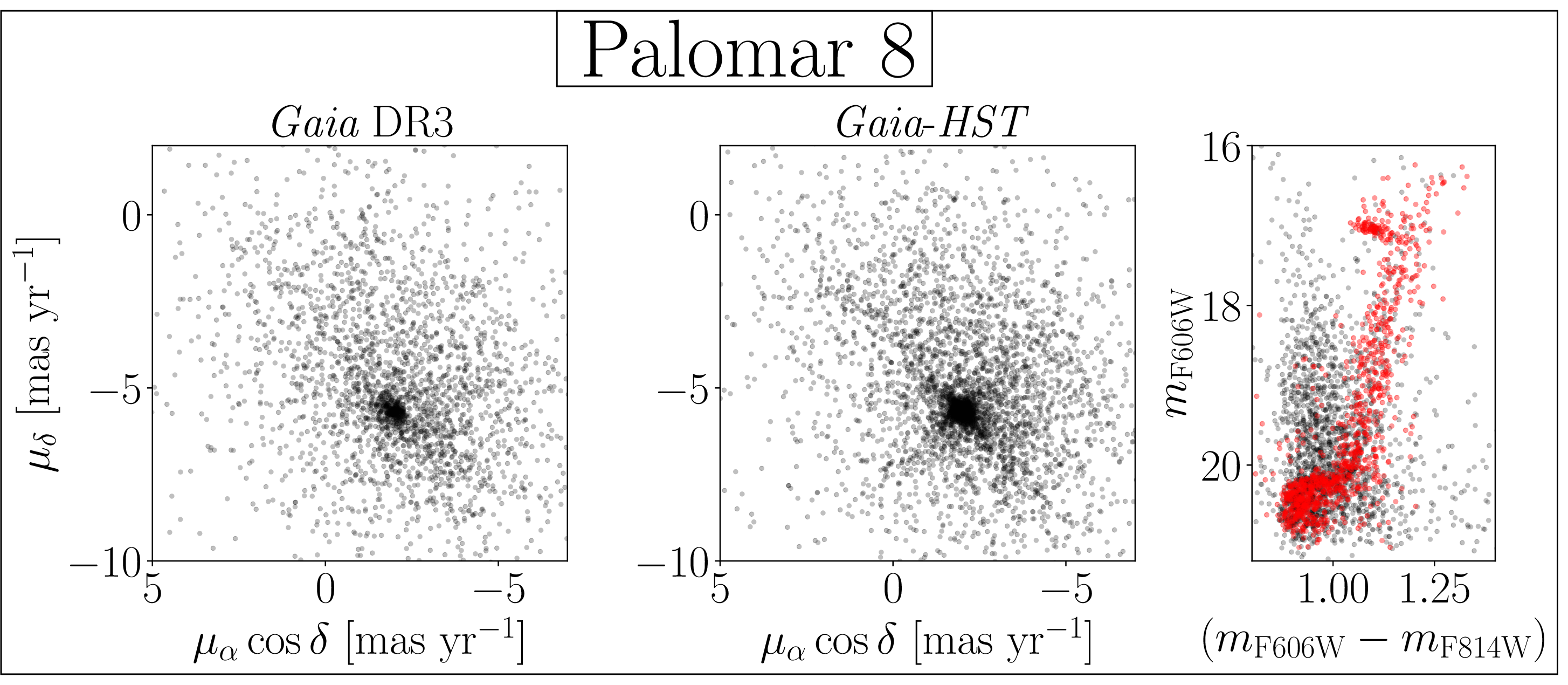}\hfill
    \includegraphics[width=9.2cm, keepaspectratio]{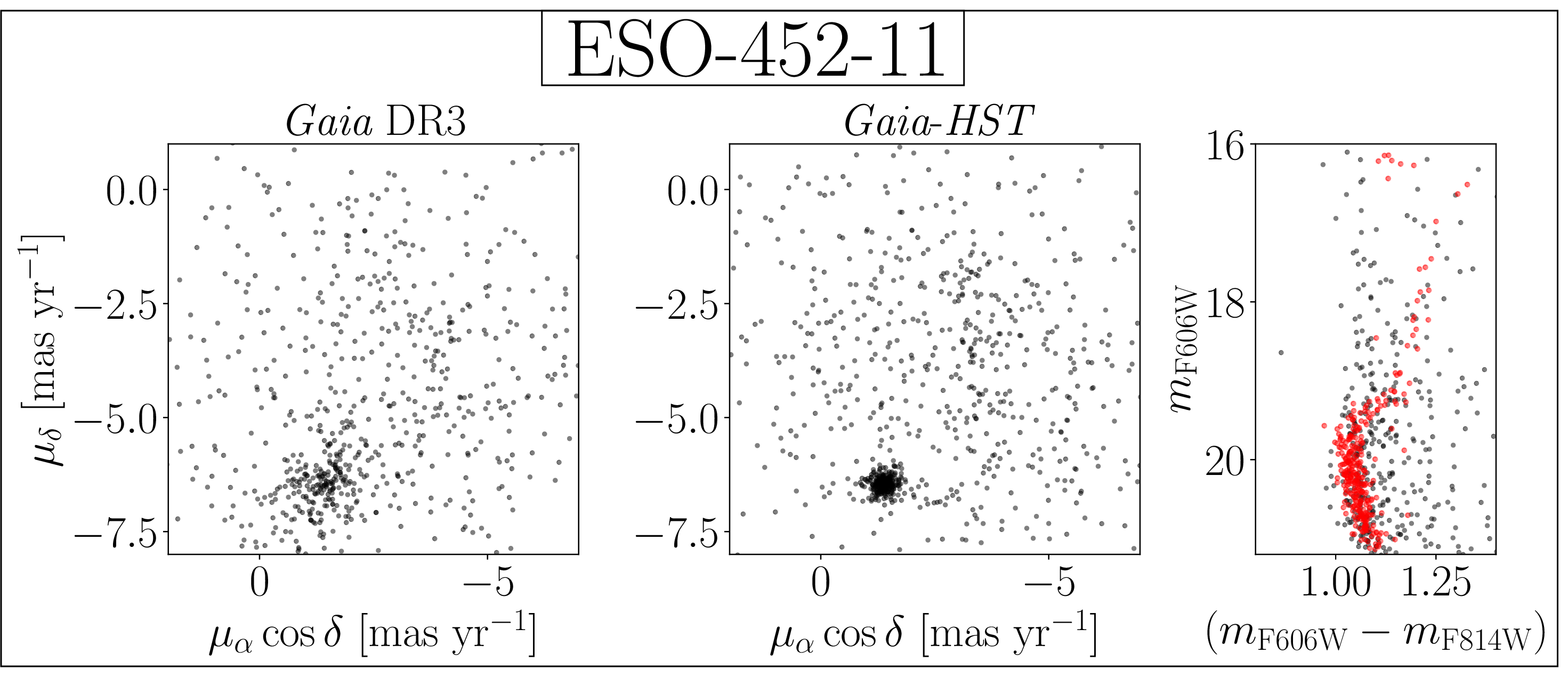}
    \includegraphics[width=9.2cm, keepaspectratio]{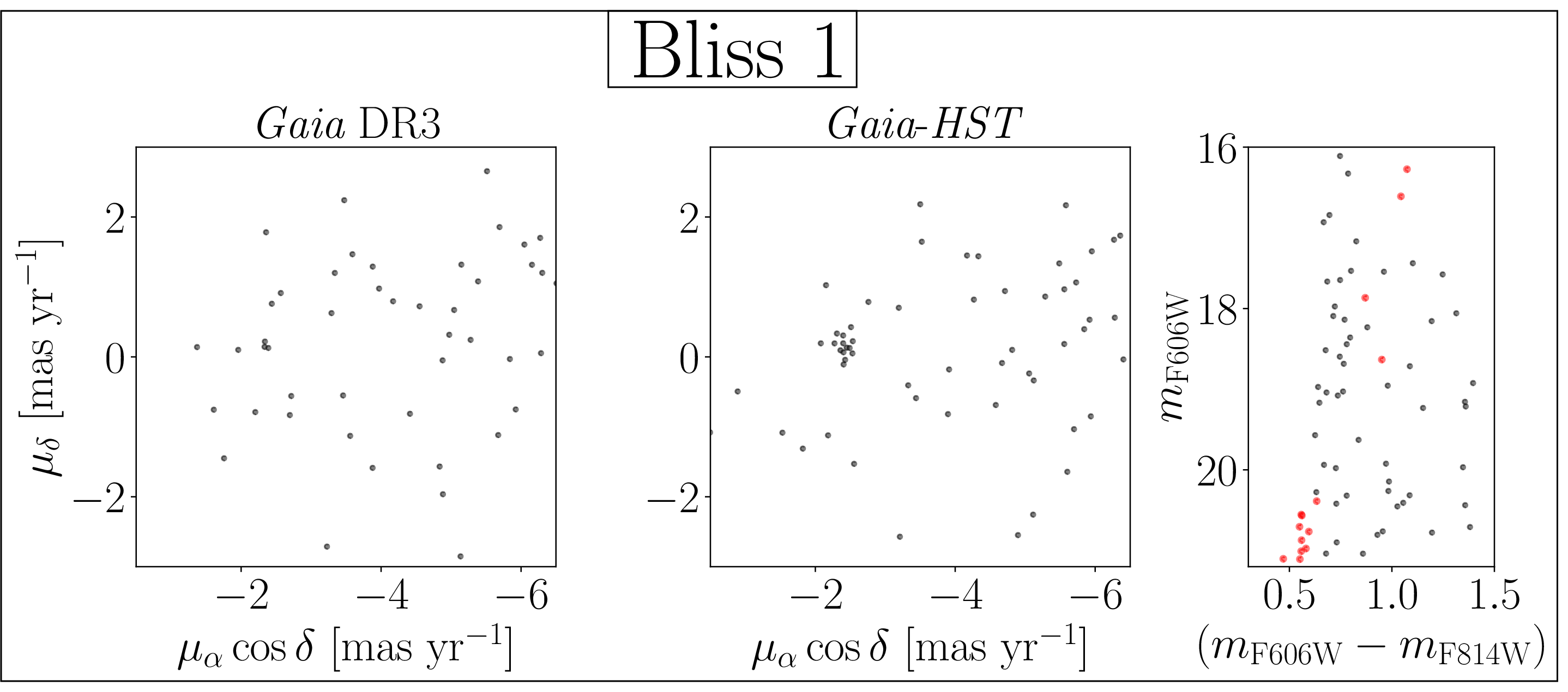}\hfill
    \includegraphics[width=9.2cm, keepaspectratio]{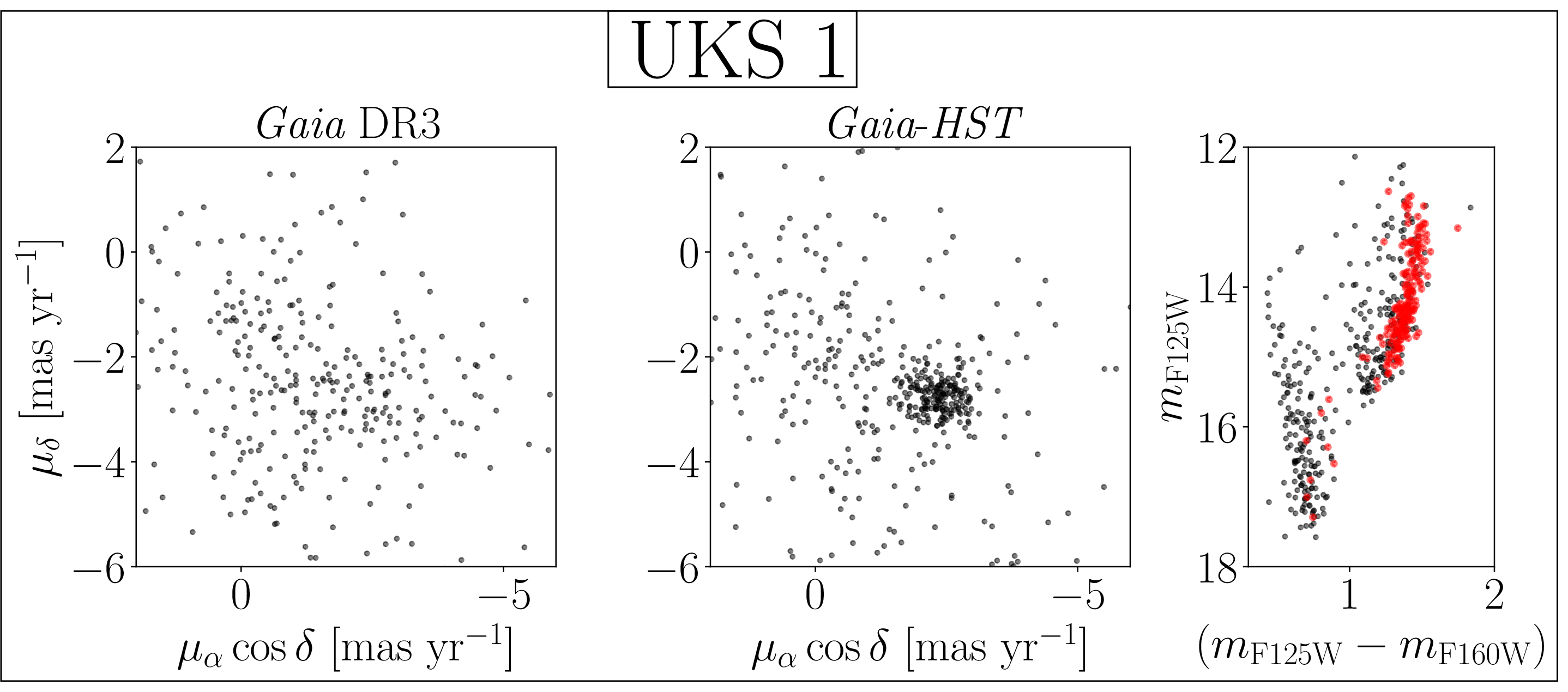}
    \caption{Absolute PMs for a sample of GCs in our program. For each GC (two for each row), from left to right we present: the VPD of the \gaia DR3 PMs; the VPD of the PMs from \texttt{GaiaHub}; and a calibrated CMD. In the rightmost CMD, stars in red are likely cluster members according to their PM (made with an arbitrary magnitude-dependent PM selection). This Figure includes examples of a very-crowded region (Palomar~8); a well-populated, moderately-crowded field (ESO-452-11); a loose GC in a sparse field (Bliss~1); and a cluster observed with the WFC3/IR camera (UKS~1).}
    \label{fig:abspm}
\end{figure*}

\section{Scientific application: Probing globular-cluster origins with \texttt{GaiaHub} proper motions}\label{sec:pm}

For most MGCS targets, cluster members in the CMD are embedded in foreground or background field stars. \citet[hereafter, Paper~II]{2026rosignoli} discuss two methods devised to facilitate the analysis of the MGCS GCs:\ a statistical approach purely based on photometry (applicable only to systems with both ACS and WFC3 optical data), and PM decontamination. As an example of a possible application, we show in the following the latter method with PMs in the central fields of our GCs computed using \texttt{GaiaHub}\footnote{We modified the publicly available code to support WFC3/IR data as well.} \citep{2022ApJ...933...76D}. \texttt{GaiaHub} is a tool designed to combine \hst and \gaia data to compute PMs. The \hst-\gaia combination has been shown to be very effective in improving the PM precision at \gaia's faint end.

We ran \texttt{GaiaHub} following the same approach as in \citetalias{2026rosignoli}. First, we bypassed the initial step of \texttt{GaiaHub} in which \hst-based astro-photometric catalogs (one for each image) are created from scratch and instead used the output of our first-pass photometry stage described in Sect.~\ref{sec:datared}. We ran \texttt{GaiaHub} with the \mbox{\texttt{-\mbox{}-rewind\_stars}} option, which uses PMs (from \gaia first, and then from \texttt{GaiaHub}) to propagate \gaia's positions at the epoch of the \hst observations to facilitate the cross-matching of the catalogs and transform positions onto the same reference system. We opted for this setup because cluster membership is not straightforward, as field stars significantly outnumber cluster stars in many cases. Due to how we run \texttt{GaiaHub}, our output PMs are already in an absolute reference frame set by the \gaia DR3 PMs. For NGC~6749, we used PMs from \citetalias{2026rosignoli}.

We ran \texttt{GaiaHub} with default settings and let it converge to the final PMs. For 29 of 34 GCs, the convergence was reached before the maximum number of iterations (ten) that the code ran. For the remaining five (Bliss~1, Gran~1, Kim~3, Koposov~1, and Mercer~5), we carefully examined the output. For Gran~1, the PM catalog presented puzzling position- and magnitude-dependent systematics that our local corrections (see below) were unable to remove. For this reason, we chose not to use the PMs for this object. For Bliss~1 and Mercer~5, we could not find any suspicious systematic in the PMs (after local corrections were applied). For the remaining two GCs (Kim~3 and Koposov~1), the few stars present in their vector-point diagrams (VPDs) of the PMs did not allow us to reach a definitive conclusion.

We also searched and corrected for systematics in the PMs that correlate with the position of stars in the field using a local correction obtained from comparing our \texttt{GaiaHub} and \gaia DR3 PMs (i.e., by subtracting for each star the median difference between our and \gaia PMs for the $N$ closest neighbors, with $10 \le N \le 50$ chosen as a compromise between the number of available sources and the need to select a network of neighbor stars as local as possible to the target to better capture the local systematics; see \citetalias{2026rosignoli}). These PM systematics can arise from inaccuracies in the \hst PSF models to residual, uncorrected \hst geometric distortion and charge-transfer-efficiency defects. It is difficult to discern the nature of these effects a posteriori from PMs, but it is straightforward to mitigate them (\citetalias{2026rosignoli}, but see also \citealt{2014BelliniPM} and \citealt{2022LibralatoPMcat}). We emphasize that even the locally-corrected PMs might not be free of systematics. The local corrections are computed starting from the assumption that the \gaia PMs, although less precise, are still accurate at the faint-end of the \gaia catalog. However, \gaia PMs are not perfect either, and in extreme regions like those targeted by our \hst observations, \gaia PMs might also have some intrinsic systematics. Therefore, any systematic in \gaia's PMs could be transferred to our PMs. In general, we advise users to carefully check the presence of systematics in the PMs before running any kinematic analysis. Demanding applications like the assessment of the internal motions could be complicated with our PMs for some GCs, especially those systems with very few stars. The raw and corrected PMs are included (when available) in our catalogs (see Appendix~\ref{appendix:release}).

We used our PMs to refine the absolute PMs of most GCs of the MGCS. For Koposov~1, Mu\~noz~1 and Segue~3, the VPDs are poorly populated and do not show any overdensity of stars linked to the clusters (not even using the current estimates of the absolute PM of these GCs provided by \citealt{2021VasilievGCkin} as a reference), but just a sparse PM distribution like that of field interlopers. 2MASS-GC01 and 2MASS-GC02 have instead moderately-populated VPDs, but we could not identify cluster members either. For these reasons, we did not compute the absolute PMs for these systems. Conversely, for Kim~3 and Koposov~2, the VPDs are not very populated, and we were unable to compute local corrections for their PMs, but stars with a motion close to that of the cluster predicted by \gaia \citep{2021VasilievGCkin} are present. We chose to include these systems in our analysis, although we are aware of the possible presence of uncorrected systematics.

The absolute PMs of \texttt{GaiaHub} were computed by computing the 3$\sigma$-clipped median value (and error) of the PMs of the cluster members (selected using both PMs and their position in the CMD). The results are summarized in Appendix~\ref{appendix:abspm}, and a few examples of our astrometric analysis are shown in Fig.~\ref{fig:abspm}. Our estimates are within 3$\sigma$ from \gaia-based values in the literature \citep[from the online repositories of GC parameters of Andrew Pace\footnote{\href{https://github.com/apace7/local_volume_database}{https://github.com/apace7/local\_volume\_database}} and Holger Baumgardt\footnote{\href{https://people.smp.uq.edu.au/HolgerBaumgardt/globular/}{https://people.smp.uq.edu.au/HolgerBaumgardt/globular/}};][]{Pace2025OJAp....8E.142P,2021VasilievGCkin}. Our uncertainties on the absolute PMs are on average 3.7 times smaller than those in the literature.

\begin{figure*}
    \sidecaption
    \centering
    \includegraphics[width=12.9cm]{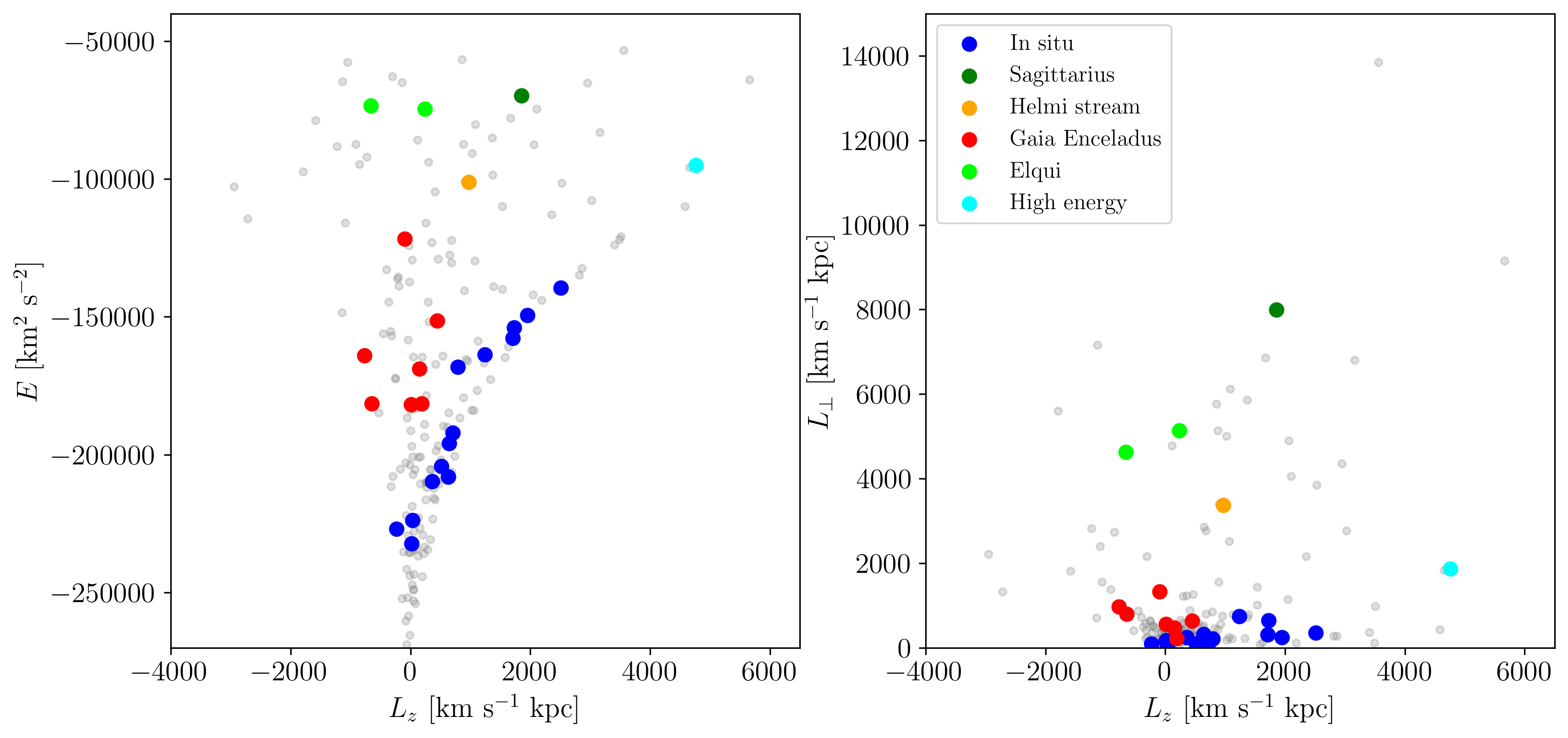}
    \caption{Projections (left panel: $E$ versus $L_z$; right panel: $L_\perp$ versus $L_z$) of the integral-of-motion space for the GCs analyzed in our work, color-coded according to the association with their progenitor (see legend in the right panel). The gray points represent all other GCs included in the repository of \citet{Pace2025OJAp....8E.142P} as a reference.}
    \label{fig:progenitors}
\end{figure*}

Given the higher precision achieved, we combined our newly-computed PMs with positions, distances and line-of-sight (LOS) velocities from \citet{Pace2025OJAp....8E.142P}\footnote{Some GCs in the catalog of \citet{Pace2025OJAp....8E.142P} do not have a measurement of either the LOS velocity or the distance error. In these cases, we used the values adopted in \citet{2025RNAAS...9...64M} for the former, and set an error equal to 50\% of the distance for the latter, respectively.} to find the associations between each GC and its putative galaxy progenitor as in \citet{2019A&A...630L...4M} and \citet{2025RNAAS...9...64M}. We used the \texttt{AGAMA} code\footnote{\href{https://github.com/GalacticDynamics-Oxford/Agama}{https://github.com/GalacticDynamics-Oxford/Agama}} \citep{2019MNRAS.482.1525V}, adopting the gravitational potential of \citet{2017MNRAS.465...76M}, to determine the following orbital parameters: total energy ($E$), vertical angular momentum ($L_z$), perpendicular angular momentum ($L_\perp$), maximum vertical height, apocenter, and circularity. We then assigned the GCs in our sample to the corresponding progenitors following the same criteria as in \citet{2025RNAAS...9...64M}. The results are shown in Fig.~\ref{fig:progenitors} and are summarized as follows:
\begin{itemize}
    \item \textit{In situ (Bulge)}: ESO-452-SC11, Gran~5, VVV-CL001;
    \item \textit{In situ (Disk)}: BH~140, BH~176, FSR~1716, Garro~1, Mercer~5, NGC~6749, Palomar~8, Palomar~10, Palomar~11, Patchick~126, PWM~2;
    \item \textit{Sagittarius}:  Whiting~1;
    \item \textit{Helmi Streams}: Gran~4;
    \item \textit{Gaia-Enceladus}: ESO-280-6, Gran~2, Gran~3, NGC~7492, RLGC-02, UKS~1, VVV-CL002;
    \item \textit{Elqui}: Laevens~3, Kim~3;
    \item \textit{High energy}: Bliss~1, Koposov~2, RLGC-01.
\end{itemize}
The progenitors identified in our work are in agreement with those found by \citet{2025RNAAS...9...64M}, with four noticeable exceptions. The first is VVV-CL001, which we now potentially associate with the \textit{in situ} Bulge rather than the accreted \textit{low-energy} group. The second is VVV-CL002, which, according to the new orbital parameters, is potentially associated with \textit{Gaia-Enceladus} rather than with the MW bulge. Finally, our results highlight the presence of two new members of \textit{Elqui}, namely Laevens~3 and Kim~3, which change their association from \textit{Sagittarius} and \textit{high-energy} groups, respectively. We note that the orbit of UKS~1 now seems more consistent with \textit{Gaia-Enceladus}, compared to its previous uncertain association. 

Our results are also consistent with those in \citet{chen24}, with the exception of UKS~1. Furthermore, a comparison with the analysis of \citet{callingham22} reveals discrepancies with our findings for approximately one third of the systems. Most clusters with different associations are in regions of the integral-of-motion space where classification is more challenging (e.g., $-20\times10^4 < E < -15\times10^4$ km$^2$ s$^{-2}$) because of the overlap of the debris from several progenitors in a limited volume of the parameter space. Furthermore, we stress that although \citet{callingham22} used metallicity and age along with dynamics to improve the identification of the progenitor, most of these GCs with discrepant associations did not have age information in their work either. Thus, we expect that our more precise PM measurements provide an improvement significant enough to make these new associations more robust. Finally, we confirm an in-situ, Disk-like origin for Patchick~126 as in \citet{2026A&A...706A.338G}.

For five GCs (Garro~1, Gran~2, Gran~3, Gran~4, Gran~5), we provide here for the first time an association based on the homogeneous dynamical framework described in \citet{2025RNAAS...9...64M}. The above assignments agree with those of \citet{deleo26} for Gran~4 and Gran~5, whereas Gran~3 is unassociated according to that work and Gran~2 is linked to a more general retrograde substructure that is included in our definition of \textit{Gaia-Enceladus}. For Garro~1 our interpretation agrees with the in situ origin inferred by \citet{pace23}. All updated associations are available on the CARMA website\footnote{\href{https://www.oas.inaf.it/en/research/m2-en/carma-en/}{https://www.oas.inaf.it/en/research/m2-en/carma-en/}} \citep{2023A&A...680A..20M}.

\section{Conclusion}

The \hst MGCS is one of the last efforts, but not least, to shed light on various aspects of MW GCs using homogeneous photometry (and astrometry). This third paper in the series focuses on the creation of the official astro-photometric catalogs of the project, which have already been used in \citetalias{2025A&A...698A.197M} and \citetalias{2026rosignoli}. All data products described in the paper are made publicly available to the community. At the time of writing, there are $\sim$170 confirmed GCs \citep[e.g.,][]{2021VasilievGCkin}. Thanks to the MGCS and other \hst programs (some of which are still ongoing) with similar setups, there will soon be 112 GCs with homogeneous ACS/WFC F606W and F814W photometry.

The \hst MGCS observing strategy secured exquisite images for 27 GCs with the ACS/WFC$+$WFC3/UVIS cameras and for seven with the WFC3/IR detector. The optical sample comprises data of both the core and the outskirts of each GC, while the IR sample includes only the core. We used state-of-the-art reduction techniques for \hst to obtain high-precision astrometry and photometry for all targets (see Appendix~\ref{appendix:cmd} for a collection of CMDs for these systems). Upon inspection of our optical catalogs, we found the presence of photometric discrepancies between the central and outer fields of the same cluster. Although we ignored the reason for this discrepancy, we devised a novel correction that takes advantage of \gaia-based synthetic photometry. This simple application shows that \gaia can serve as a powerful auxiliary catalog for photometry, analogously to what is currently done with astrometry.

Our work also includes comprehensive artificial-star tests for each GC. The resulting ancillary catalogs serve a dual purpose: they enable a precise assessment of the completeness level and provide a robust framework for investigating and characterizing systematic errors in our data.

We combined data from \hst and \gaia using \texttt{GaiaHub} to determine precise PMs for the MGCS targets. Although the magnitude interval (typically 4--5 magnitudes) covered by both \gaia and \hst is not enough to fully exploit our \hst data, the synergy between \gaia and \hst makes it possible to identify cluster members in systems where \gaia alone struggles (see also \citetalias{2026rosignoli}). These PMs allowed us to refine the absolute PMs of the clusters, finding estimates that are, on average, 3.7 times more precise than those in the literature based on the \gaia DR3 catalog. 

We integrated our \hst-\gaia PMs with positions, distances, and LOS velocities derived from the literature to characterize the clusters’ orbital parameters. This allowed us to update--and in several cases, identify for the first time--associations between these GCs and their putative progenitor galaxies, ultimately contributing to the reconstruction of the MW’s assembly history.

We release astro-photometric catalogs, astrometrized stacked images, and artificial-star tests to the community. These catalogs increase the sample of GCs with exquisite photometry, enabling the scientific community to further investigate the formation, dynamics, and evolutionary history of the MW and its satellites. These comprehensive data continue the legacy of \hst, and provide a bridge for future surveys with \jwst and \textit{Nancy Grace Roman} Space Telescope.

\section*{Data availability} Astro-photometric catalogs and stacked images are available at the CDS via anonymous ftp to to \href{cdsarc.u-strasbg.fr}{cdsarc.u-strasbg.fr} (130.79.128.5) or via \href{https://cdsarc.cds.unistra.fr/viz-bin/cat/J/A+A/709/A140}{https://cdsarc.cds.unistra.fr/viz-bin/cat/J/A$+$A/709/A140}, and at our website \href{https://www.oas.inaf.it/en/research/m2-en/mgcs-en/}{https://www.oas.inaf.it/en/research/m2-en/mgcs-en/}.

\begin{acknowledgements}
DM, SC, EP and AM acknowledge financial support from PRIN-MIUR-22: CHRONOS: adjusting the clock(s) to unveil the CHRONO-chemo-dynamical Structure of the Galaxy” (PI: S. Cassisi). SC acknowledges the support of a fellowship from La Caixa Foundation (ID 100010434) with fellowship code LCF/BQ/PI23/11970031 (P.I.: A. Escorza) and from the Fundación Occident and the Instituto de Astrofísica de Canarias under the  Visiting Researcher Programme 2022-2025 agreed between both institutions. SS acknowledges funding from the European Union under the grant ERC-2022-AdG, ``StarDance: the non-canonical evolution of stars in clusters'', Grant Agreement 101093572, PI: E. Pancino. RP acknowledges the support to this study by the INAF Mini Grant 2025 (Ob.Fu.1.05.24.07.05, CUP C33C24001390005). Based on observations with the NASA/ESA \textit{HST}, obtained at the Space Telescope Science Institute, which is operated by AURA, Inc., under NASA contract NAS 5-26555. Support for Program number GO-17435 was provided through grants from STScI under NASA contract NAS5-26555. This work has made use of data from the European Space Agency (ESA) mission {\it Gaia} (\url{https://www.cosmos.esa.int/gaia}), processed by the {\it Gaia} Data Processing and Analysis Consortium (DPAC, \url{https://www.cosmos.esa.int/web/gaia/dpac/consortium}). Funding for the DPAC has been provided by national institutions, in particular the institutions participating in the {\it Gaia} Multilateral Agreement. This job has made use of the Python package GaiaXPy, developed and maintained by members of the Gaia Data Processing and Analysis Consortium (DPAC), and in particular, Coordination Unit 5 (CU5), and the Data Processing Centre located at the Institute of Astronomy, Cambridge, UK (DPCI). This research made use of \texttt{astropy}, a community-developed core \texttt{python} package for Astronomy \citep{astropy:2013, astropy:2018}.
\end{acknowledgements}

\bibliographystyle{aa}
\bibliography{bibliography}

\begin{appendix}

\section{Description of the astro-photometric catalogs}\label{appendix:release}

The content of the astrometric and photometric catalogs released to the community is described in Tables~\ref{tab:pmcat} and \ref{tab:photcat}. The astrometric catalogs of Gran~1, Koposov~1, Mu\~noz~1, Segue~3, 2MASS-GC01 and 2MASS-GC02 do not include PMs because we could not find any cluster member in the VPD, or PMs did not pass our tests (see Sect.~\ref{sec:pm}). For Kim~3 and Koposov~2, no PM corrections were computed, so raw and corrected PMs are identical. The artificial-star-test input-output catalogs are described in Tables~\ref{tab:astroart} and \ref{tab:photart}.

\begin{table*}[!b]
  \caption{Description of an astrometric catalog.}
  \centering
  \label{tab:pmcat}
  \begin{tabular}{cccl}
    \hline
    \hline
    Column & Name & Unit & Description \\
    \hline
    1  & R.A. & deg & Right Ascension \\
    2  & Dec. & deg & Declination \\
    3  & $x$ & pixel & $x$ master-frame position \\
    4  & $y$ & pixel & $y$ master-frame position \\
    5  & ID  &  & ID number of the source \\
    6  & $(\mu_\alpha \cos\delta)_{\rm raw}$ & mas yr$^{-1}$ & \hst-\gaia raw PM along $\alpha \cos\delta$ \\
    7  & $(\sigma_{\mu_\alpha \cos\delta})_{\rm raw}$ & mas yr$^{-1}$ & Error on the \hst-\gaia raw PM along $\alpha \cos\delta$ \\
    8  & $(\mu_\delta)_{\rm raw}$ & mas yr$^{-1}$ & \hst-\gaia raw PM along $\delta$ \\
    9  & $(\sigma_{\mu_\delta})_{\rm raw}$ & mas yr$^{-1}$ & Error on the \hst-\gaia raw PM along $\delta$ \\
    10 & $(\mu_\alpha \cos\delta)_{\rm corr}$ & mas yr$^{-1}$ & \hst-\gaia corrected PM along $\alpha \cos\delta$ \\
    11 & $(\sigma_{\mu_\alpha \cos\delta})_{\rm corr}$ & mas yr$^{-1}$ & Error on the \hst-\gaia corrected PM along $\alpha \cos\delta$ \\
    12 & $(\mu_\delta)_{\rm corr}$ & mas yr$^{-1}$ & \hst-\gaia corrected PM along $\delta$ \\
    13 & $(\sigma_{\mu_\delta})_{\rm corr}$ & mas yr$^{-1}$ & Error on the \hst-\gaia corrected PM along $\delta$ \\
    14 & SOURCE\_ID & & \texttt{SOURCE\_ID} from the \gaia DR3 catalog \\
    \hline
  \end{tabular}
  \tablefoot{(i) The coordinates of the center of the cluster used as a reference are provided in the header of the catalog. (ii) The pixel scale is provided in the header of the catalog. (iii) Columns from 6 to 14 are included in the catalog only if PMs were computed using \texttt{GaiaHub} (see Sect.~\ref{sec:pm}).}
\end{table*}

\begin{table*}[!b]
  \caption{Description of a photometric catalog for one filter and camera.}
  \centering
  \label{tab:photcat}
  \begin{tabular}{ccl}
    \hline
    \hline
    Column & Name & Description \\
    \hline
    1  & sat & Saturation flag \\
    2  & $m$ & Calibrated VEGA magnitude \\
    3  & $\sigma_m$ & Photometric rms \\
    4  & \texttt{QFIT} & Quality-of-PSF-fit (QFIT) parameter \\
    5  & $o$ & Fractional flux within the fitting radius prior to neighbor subtraction \\
    6  & $N_{\rm f}$ & Number of exposures in which a source was found \\
    7  & $N_{\rm u}$ & Number of exposures used to measure the flux of a source \\
    8  & \texttt{RADXS} & Excess/deficiency of flux outside the core of the star \\
    9  & sky & Sky in electrons (optical data) or electron s$^{-1}$ (IR data) \\
    10 & $\sigma_{\rm sky}$ & Sky rms in electrons (optical data) or electron s$^{-1}$ (IR data) \\
    11 & ID & ID number of the source \\
    \hline
  \end{tabular}
  \tablefoot{(i) The ID number is the same as in the corresponding astrometric catalog. (ii) All values for a source are set to 0 if the star is not measured in that filter and camera. (iii) All values but the calibrated magnitude are set to 0 if a source is saturated in the catalog. (iv) The saturation flag is set to 0 for a measurement obtained from a long exposure, 1 for a measurement obtained from a short exposure, and 9 if the star is saturated (or unsaturated but missed by the second-pass-photometry tool, i.e., likely a spurious detection only included for completeness; see Sect.~\ref{sec:datared}). (v) To estimate the significance of a source over the sky, first convert calibrated VEGA magnitudes into instrumental fluxes (in unit of electrons or electron s$^{-1}$ for optical and IR data, respectively): flux $=$ $10^{-0.4({\rm mag} - {\rm ZP_{Vega-mag}})}$. The VEGA-magnitude zero-point is provided in the header of the catalog.}
\end{table*}

\begin{table*}[t!]
  \caption{Description of an artificial-star astrometric catalog.}
  \centering
  \label{tab:astroart}
  \begin{tabular}{cccl}
    \hline
    \hline
    Column & Name & Unit & Description \\
    \hline
    1  & (R.A.)$^{\rm input}$ & deg & Input Right Ascension\\
    2  & (Dec.)$^{\rm input}$ & deg & Input Declination\\
    3  & $x^{\rm input}$ & pixel & Input $x$ master-frame position\\
    4  & $y^{\rm input}$ & pixel & Input $y$ master-frame position\\
    5  & (R.A.)$^{\rm output}$ & deg & Output Right Ascension\\
    6  & (Dec.)$^{\rm output}$ & deg & Output Declination\\
    7  & $x^{\rm output}$ & pixel & Output $x$ master-frame position\\
    8  & $y^{\rm output}$ & pixel & Output $y$ master-frame position\\
    9  & ID & & ID number of the source \\
    \hline
  \end{tabular}
  \tablefoot{(i) The coordinates of the center of the cluster and the pixel scale are the same as in the analog astrometric catalog with real sources.}
\end{table*}

\begin{table*}[t!]
  \caption{Description of an artificial-star photometric catalog for one filter and camera.}
  \centering
  \label{tab:photart}
  \begin{tabular}{ccl}
    \hline
    \hline
    Column & Name & Description \\
    \hline
    1  & $m^{\rm input}$ & Input calibrated VEGA magnitude \\
    2  & $m^{\rm output}$ & Output calibrated VEGA magnitude \\
    3  & $\sigma_m^{\rm output}$ & Output photometric rms \\
    4  & \texttt{QFIT}$^{\rm output}$ & Output quality-of-PSF-fit (QFIT) parameter \\
    5  & $o^{\rm output}$ & Output fractional flux within the fitting radius prior to neighbor subtraction \\
    6  & $N_{\rm f}^{\rm output}$ & Output number of exposures in which a source was found \\
    7  & $N_{\rm u}^{\rm output}$ & Output number of exposures used to measure the flux of a source \\
    8  & \texttt{RADXS}$^{\rm output}$ & Output excess/deficiency of flux outside the core of the star \\
    9  & sky$^{\rm output}$ & Output sky in electrons (optical data) or electron s$^{-1}$ (IR data) \\
    10 & $\sigma_{\rm sky}^{\rm output}$ & Output sky rms in electrons (optical data) or electron s$^{-1}$ (IR data) \\
    11 & ID & ID number of the source \\
    \hline
  \end{tabular}
  \tablefoot{(i) The ID number is the same as in the corresponding artificial-star astrometric catalog. (ii) All values for a source are set to 0 if the star is not measured in that filter and camera. (iii) The VEGA-magnitude zero-point is provided in the header of the catalog and it is the same as for the analog photometric catalog with real stars.}
\end{table*}

\section{Absolute proper motions}\label{appendix:abspm}

Table~\ref{tab:abspm} provides the absolute PMs of the GCs in the MGCS for which this measurement was feasible (enough stars available, clear understanding of the systematics).

\begin{table*}[t!]
    \centering
    \caption{Absolute PMs of GCs for which such measurement was feasible.}
    \label{tab:abspm}
    \begin{tabular}{cc|cc}
    \hline
    \hline
    Cluster & $ (\mu_\alpha \cos\delta, \mu_\delta)$ $[$mas yr$^{-1}$$]$ & Cluster & $(\mu_\alpha \cos\delta, \mu_\delta)$ $[$mas yr$^{-1}$$]$ \\
    \hline
    \multicolumn{4}{c}{ACS/WFC fields}\\
    \hline
	BH~140             & $(-14.853 \pm 0.004 ,  1.242 \pm 0.004)$ & Gran~4             & $(  0.475 \pm 0.017 , -3.538 \pm 0.017)$ \\
	BH~176             & $( -3.966 \pm 0.009 , -3.080 \pm 0.009)$ & Gran~5             & $( -5.490 \pm 0.034 , -9.335 \pm 0.026)$ \\
    Bliss 1            & $( -2.404 \pm 0.037 ,  0.111 \pm 0.051)$ & Kim~3\tablefootmark{$\ast$}       & $( -0.950 \pm 0.053 ,  3.831 \pm 0.030)$ \\
    ESO-280-6          & $( -0.721 \pm 0.015 , -2.826 \pm 0.012)$ & Koposov~2\tablefootmark{$\ast$}   & $( -0.745 \pm 0.056 ,  0.140 \pm 0.042)$ \\
    ESO-452-11         & $( -1.381 \pm 0.009 , -6.458 \pm 0.009)$ & Laevens~3          & $(  0.317 \pm 0.026 , -0.557 \pm 0.025)$ \\
    Palomar~8          & $( -1.991 \pm 0.008 , -5.693 \pm 0.007)$ & NGC~6749\tablefootmark{$\dagger$} & $( -2.820 \pm 0.008 , -5.990 \pm 0.008)$ \\
    FSR~1716           & $( -4.343 \pm 0.013 , -8.878 \pm 0.012)$ & NGC~7492           & $(  0.768 \pm 0.011 , -2.330 \pm 0.009)$ \\
    Garro~1            & $( -4.431 \pm 0.015 , -1.127 \pm 0.015)$ & Patchick~126       & $( -4.919 \pm 0.018 , -6.931 \pm 0.019)$ \\
    Palomar~10         & $( -4.325 \pm 0.007 , -7.153 \pm 0.007)$ & PWM~2              & $( -2.722 \pm 0.018 , -4.190 \pm 0.018)$ \\
    Palomar~11         & $( -1.770 \pm 0.009 , -4.955 \pm 0.010)$ & RLGC-01            & $(  0.971 \pm 0.017 ,  0.761 \pm 0.014)$ \\
    Gran~2             & $(  0.173 \pm 0.018 , -2.518 \pm 0.016)$ & Whiting~1          & $( -0.414 \pm 0.028 , -2.183 \pm 0.037)$ \\
    Gran~3             & $( -3.813 \pm 0.014 ,  0.688 \pm 0.012)$ \\
    \hline
    \multicolumn{4}{c}{WFC3/IR fields}\\
    \hline
	Mercer-5  & $( -3.779 \pm 0.043 , -7.285 \pm 0.064)$ & VVV-CL001 & $( -3.285 \pm 0.039 , -1.745 \pm 0.031)$ \\
	RLGC-02   & $( -2.315 \pm 0.033 , -1.864 \pm 0.027)$ & VVV-CL002 & $( -9.145 \pm 0.032 ,  2.347\pm 0.040)$ \\
	UKS~1     & $( -2.374 \pm 0.028 , -2.675 \pm 0.023)$ \\
    \hline
    \end{tabular}
    \tablefoot{
    \tablefoottext{$\ast$}{Obtained using raw, uncorrected PMs (because a robust PM correction was not possible). The PM estimates for these clusters are made with small statistics.}
    \tablefoottext{$\dagger$}{From \citetalias{2026rosignoli}.}
    }
\end{table*}

\section{Color-magnitude diagrams}\label{appendix:cmd}

\begin{figure*}
    \centering
    \includegraphics[height=3.6 cm]{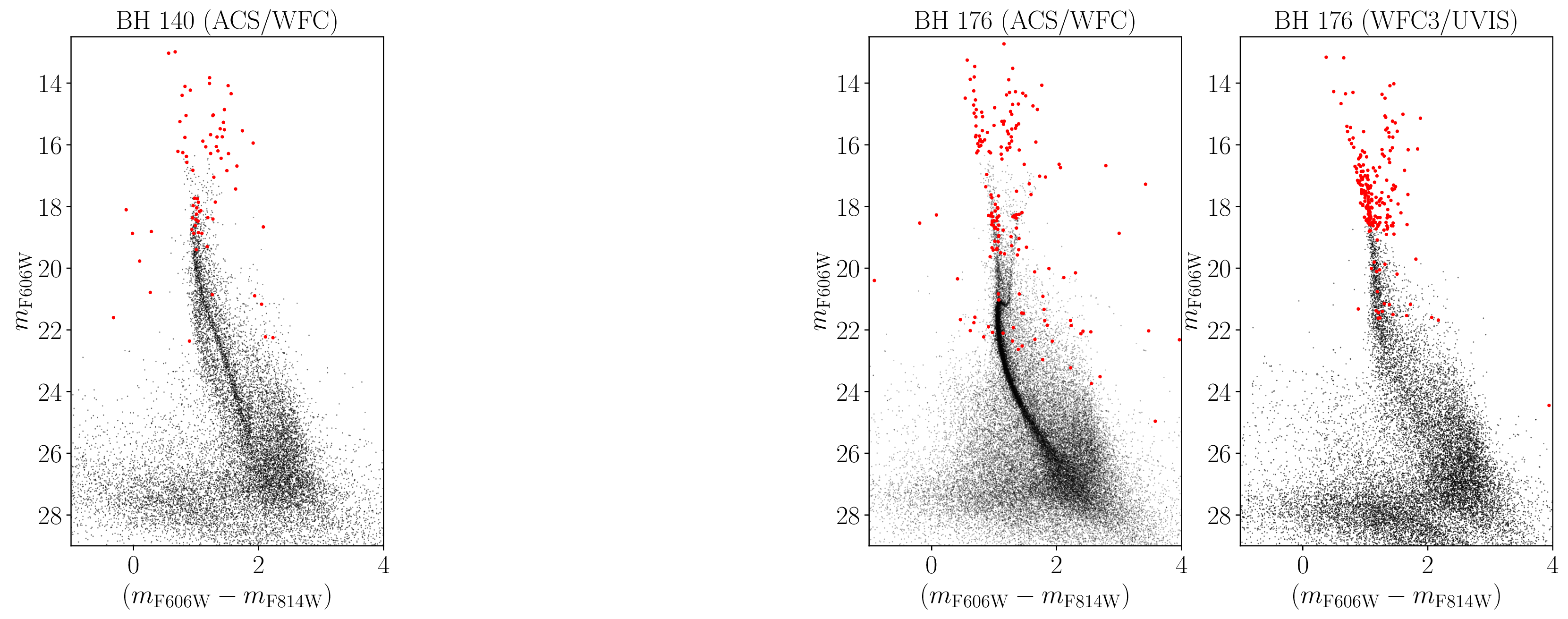}
    \includegraphics[height=3.6 cm]{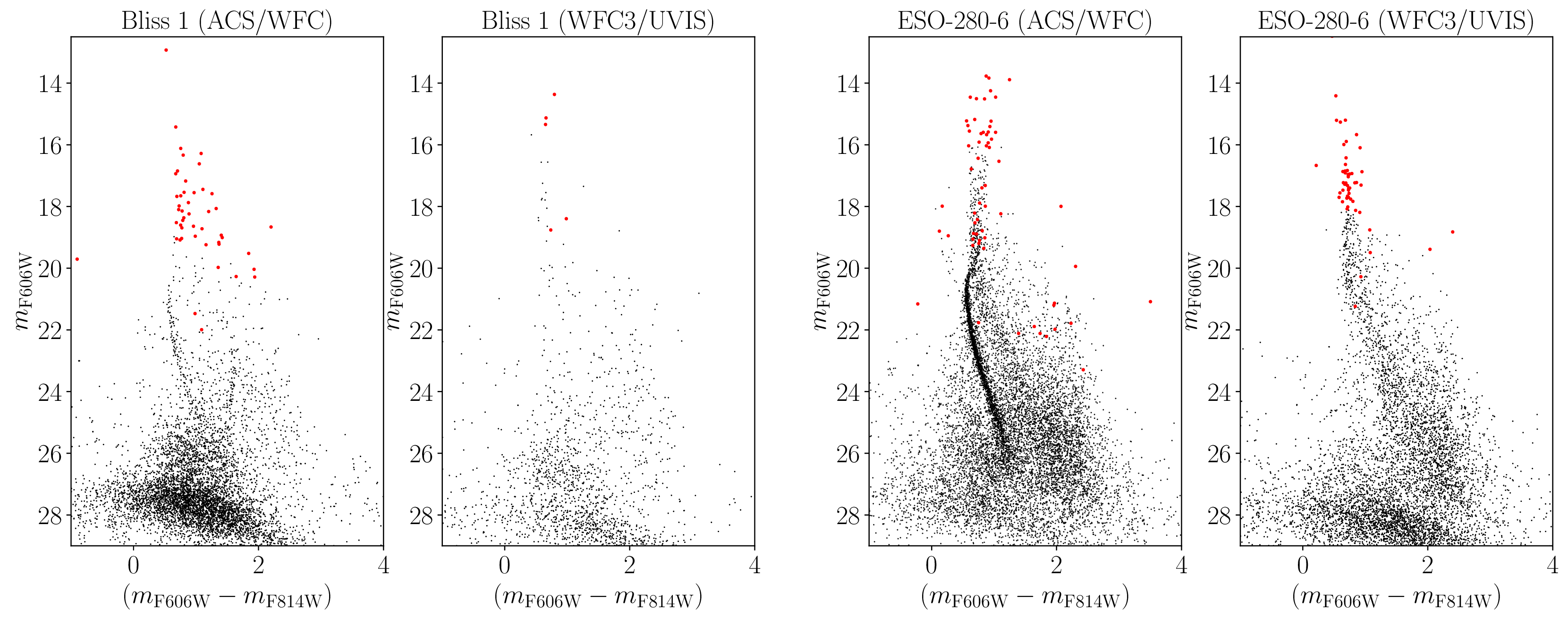}
    \includegraphics[height=3.6 cm]{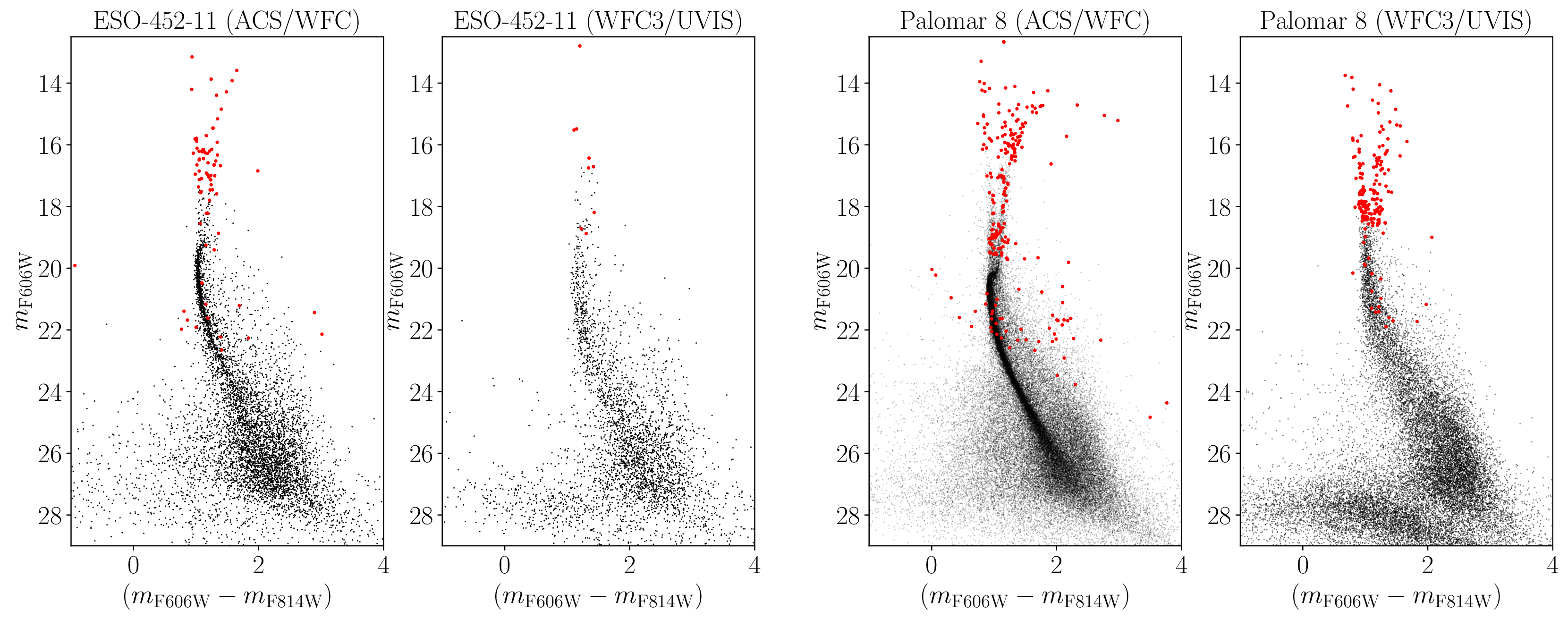}
    \includegraphics[height=3.6 cm]{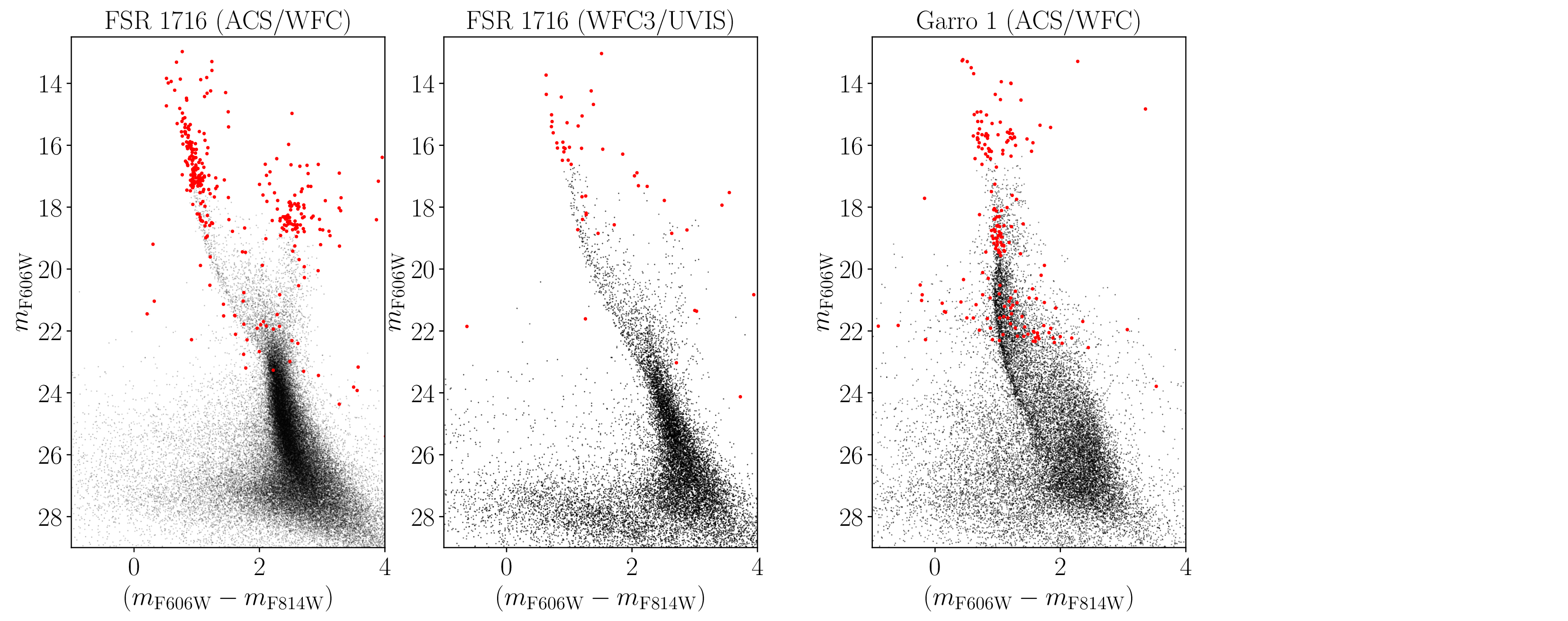}
    \includegraphics[height=3.6 cm]{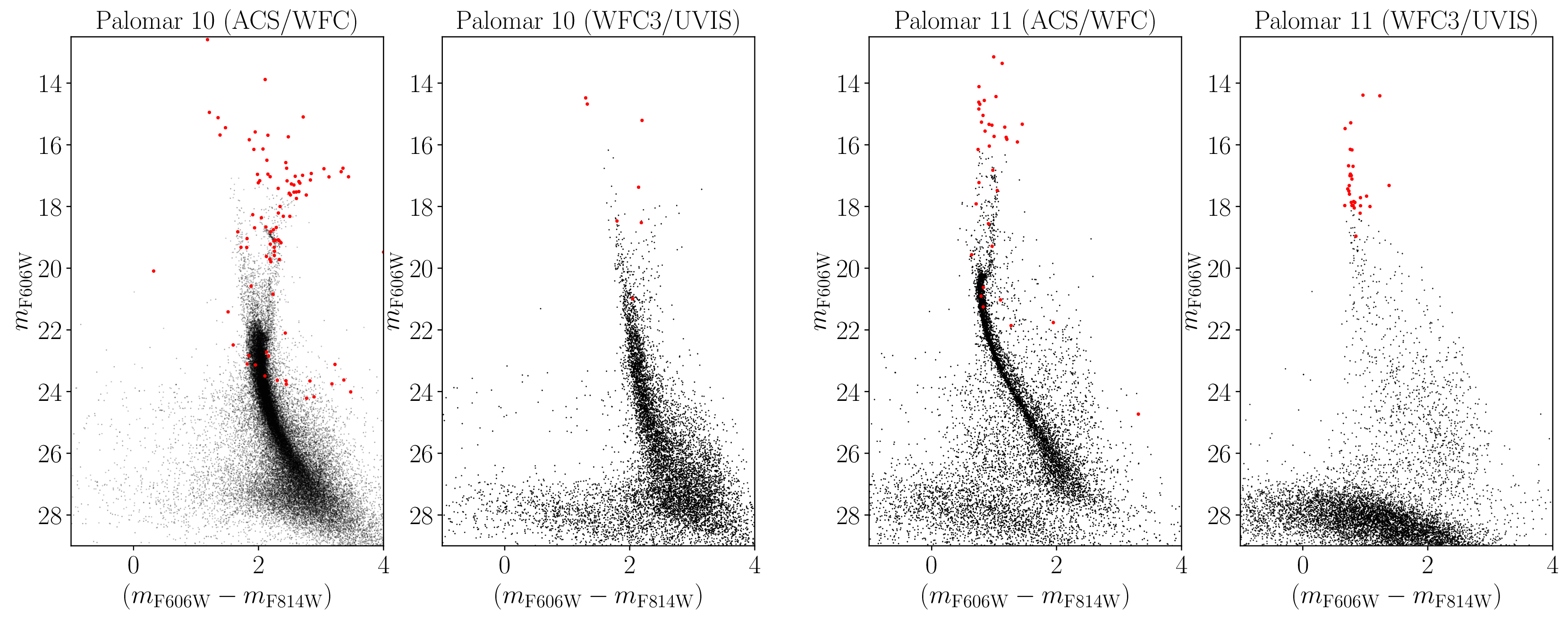}
    \includegraphics[height=3.6 cm]{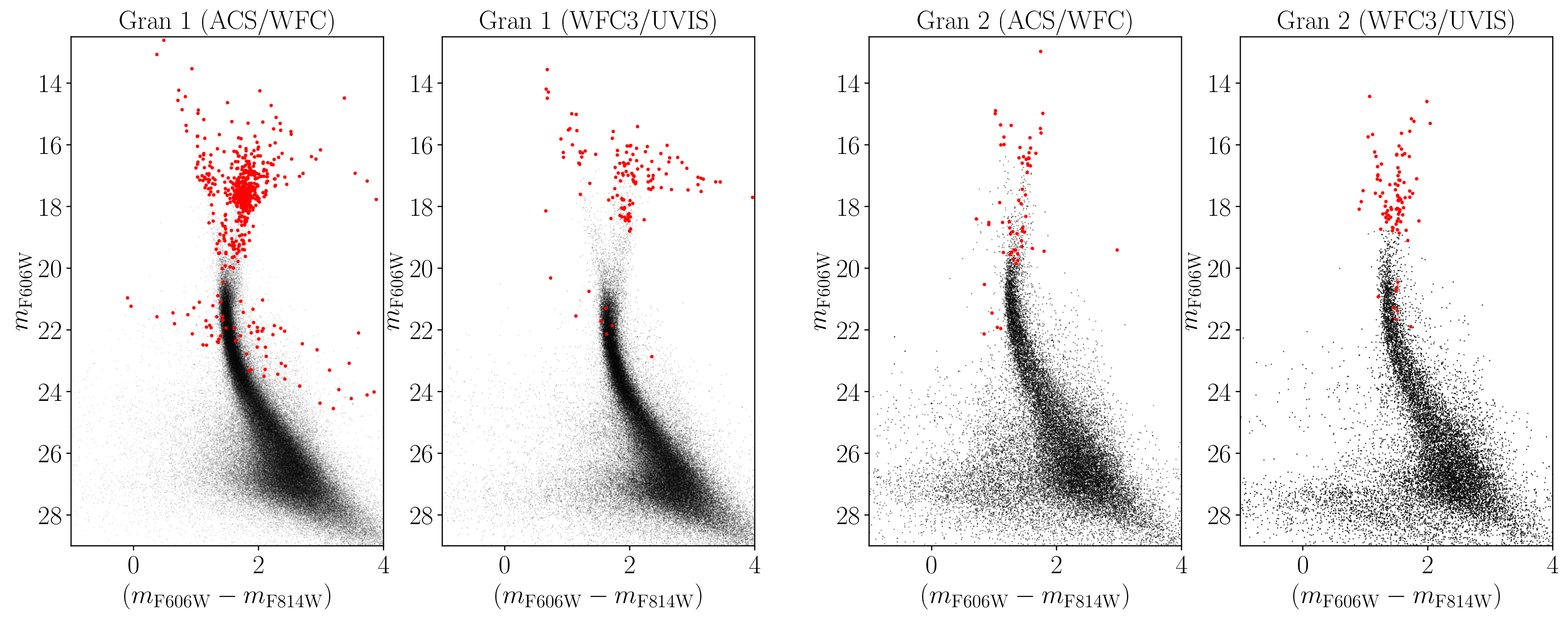}
    \caption{Collection of optical CMDs for BH-140, BH-176, Bliss\,1, ESO-280-6, ESO-452-11, Palomar~8, FSR~1716, Garro~1, Palomar~10, Palomar~11, Gran~1, and Gran~2. For each cluster, the left panel presents the ACS/WFC-based CMD of the central field, whereas the right panel (when present) shows the WFC3/UVIS-based CMD of the parallel field. Red points represent stars saturated (or unsaturated but missed by \texttt{KS2}; see discussion in Sect.~\ref{sec:datared}) in either filter. All other sources are shown as black dots.}
    \label{fig:cmd0}
\end{figure*}

\begin{figure*}
    \centering
    \includegraphics[height=3.6 cm]{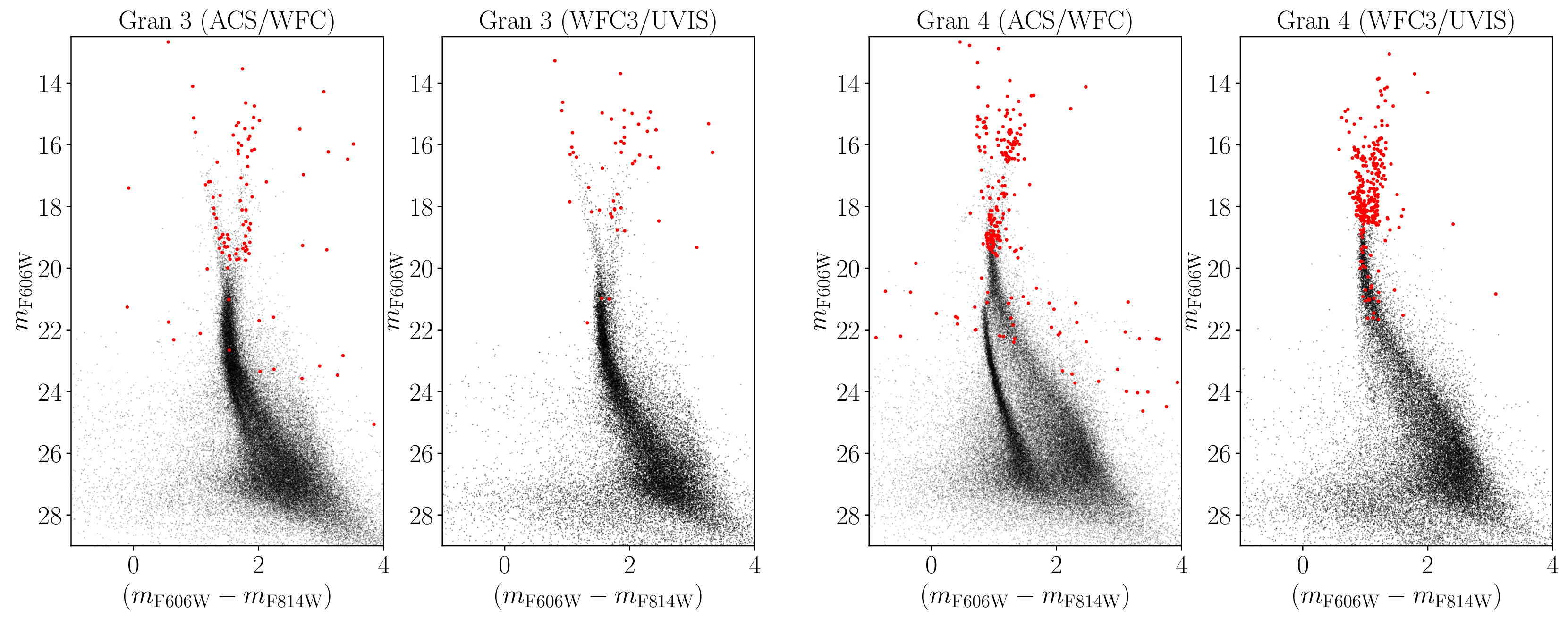}
    \includegraphics[height=3.6 cm]{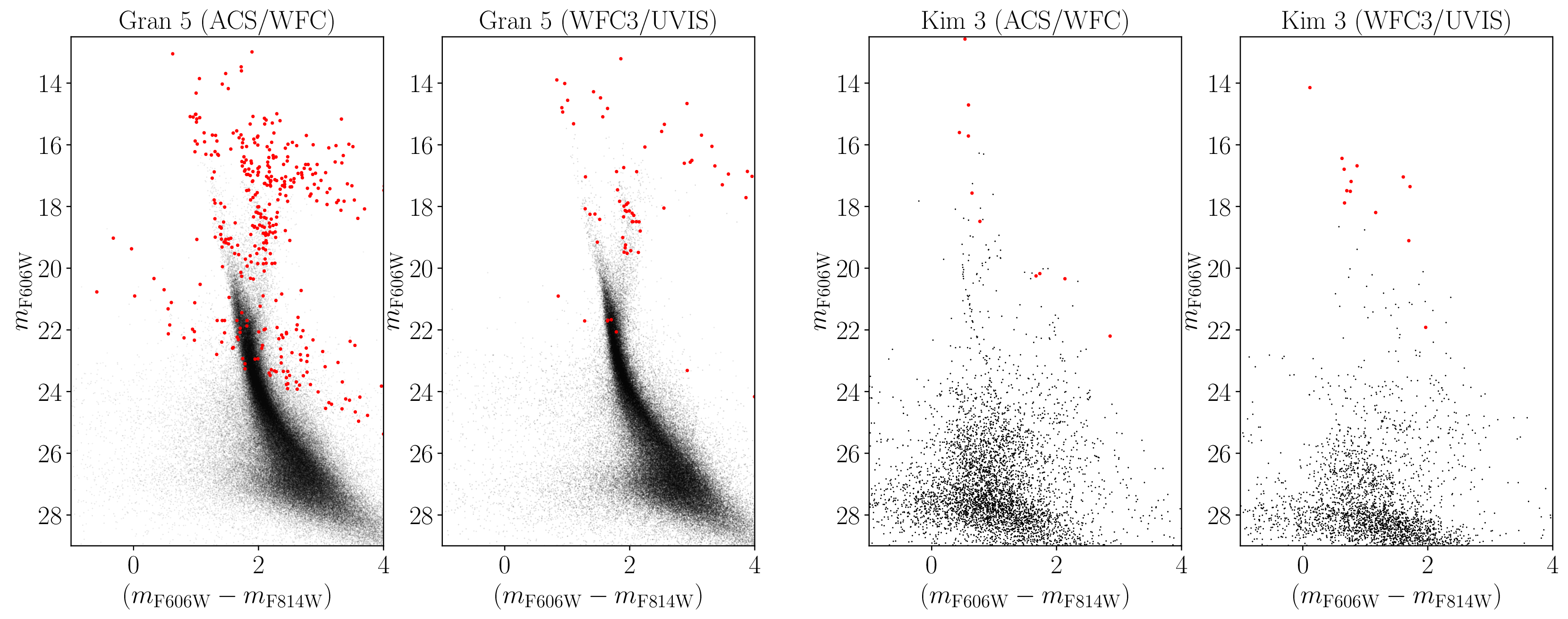}
    \includegraphics[height=3.6 cm]{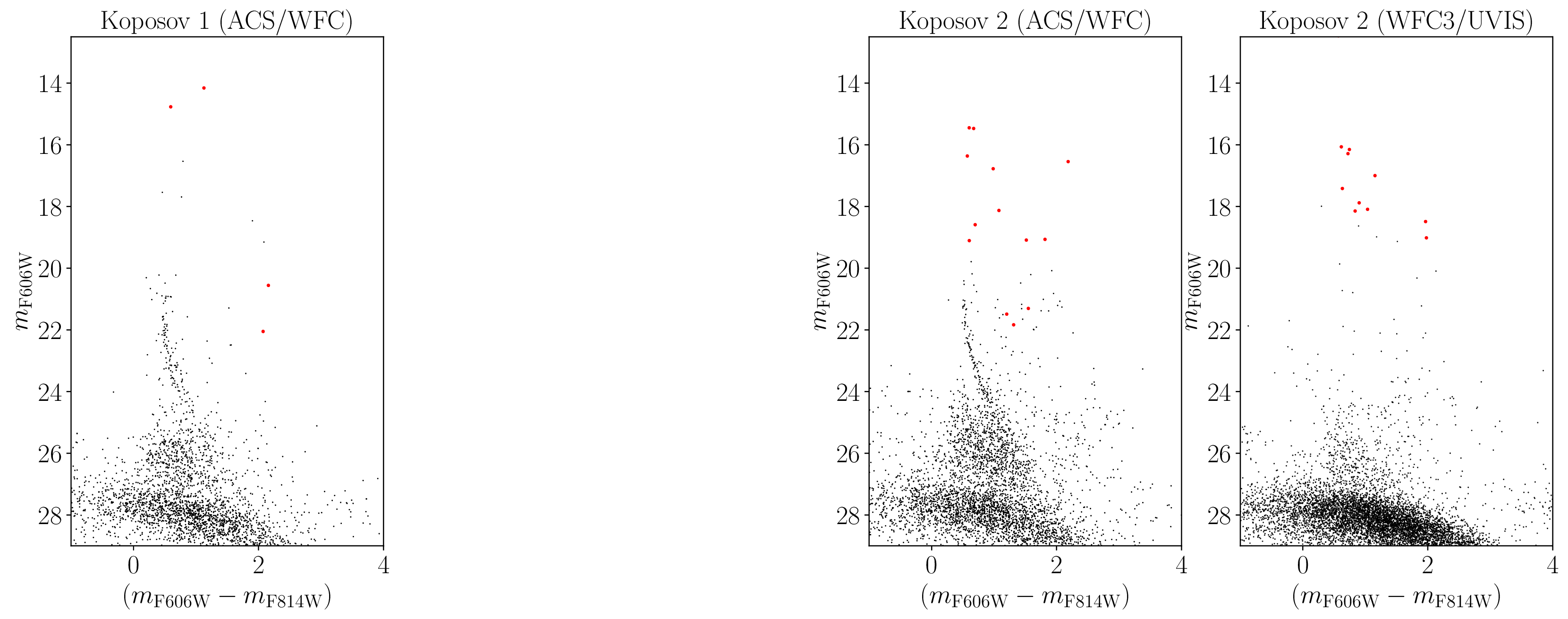}
    \includegraphics[height=3.6 cm]{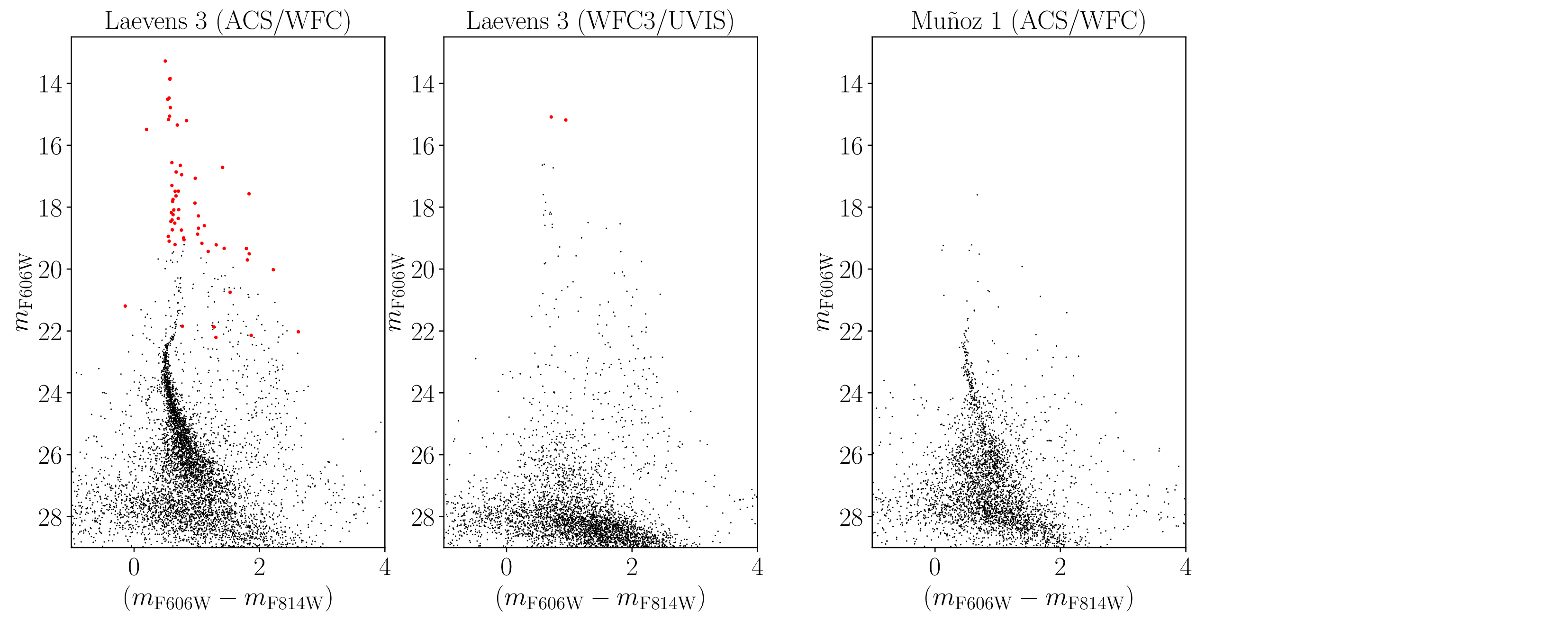}
    \includegraphics[height=3.6 cm]{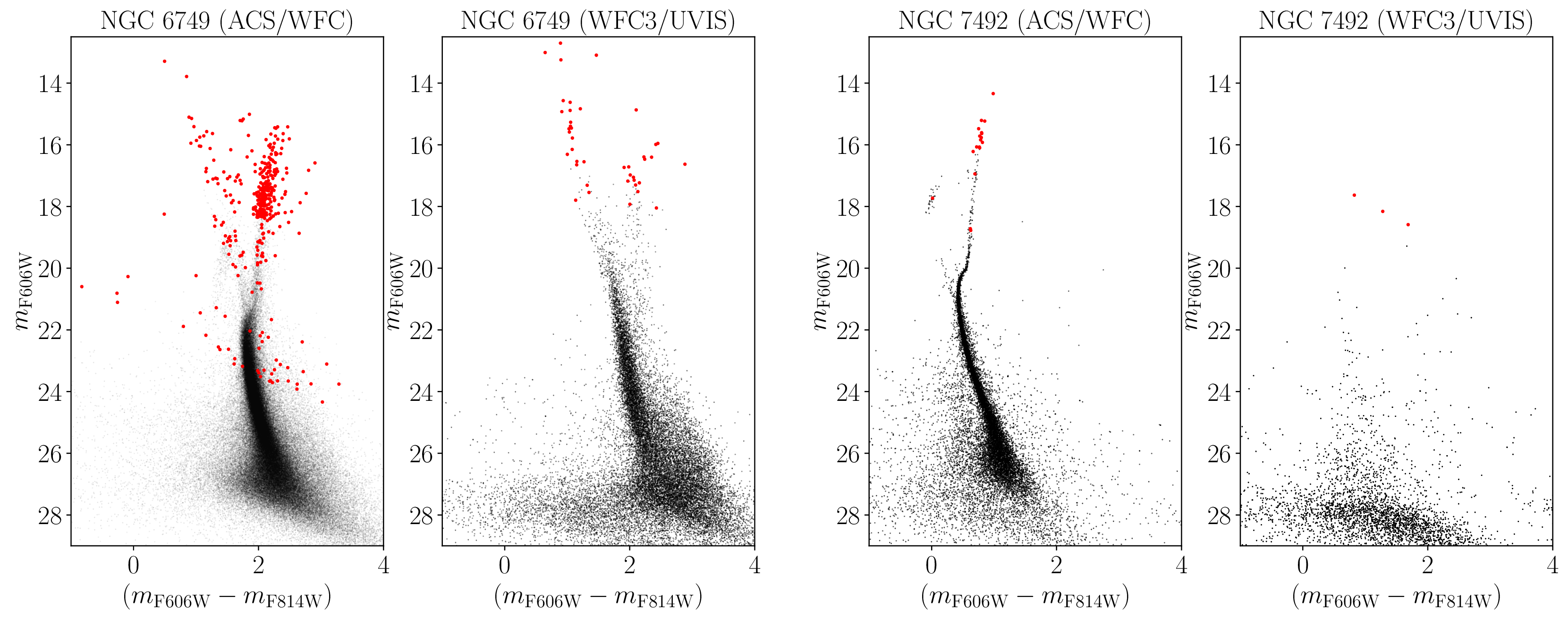}
    \includegraphics[height=3.6 cm]{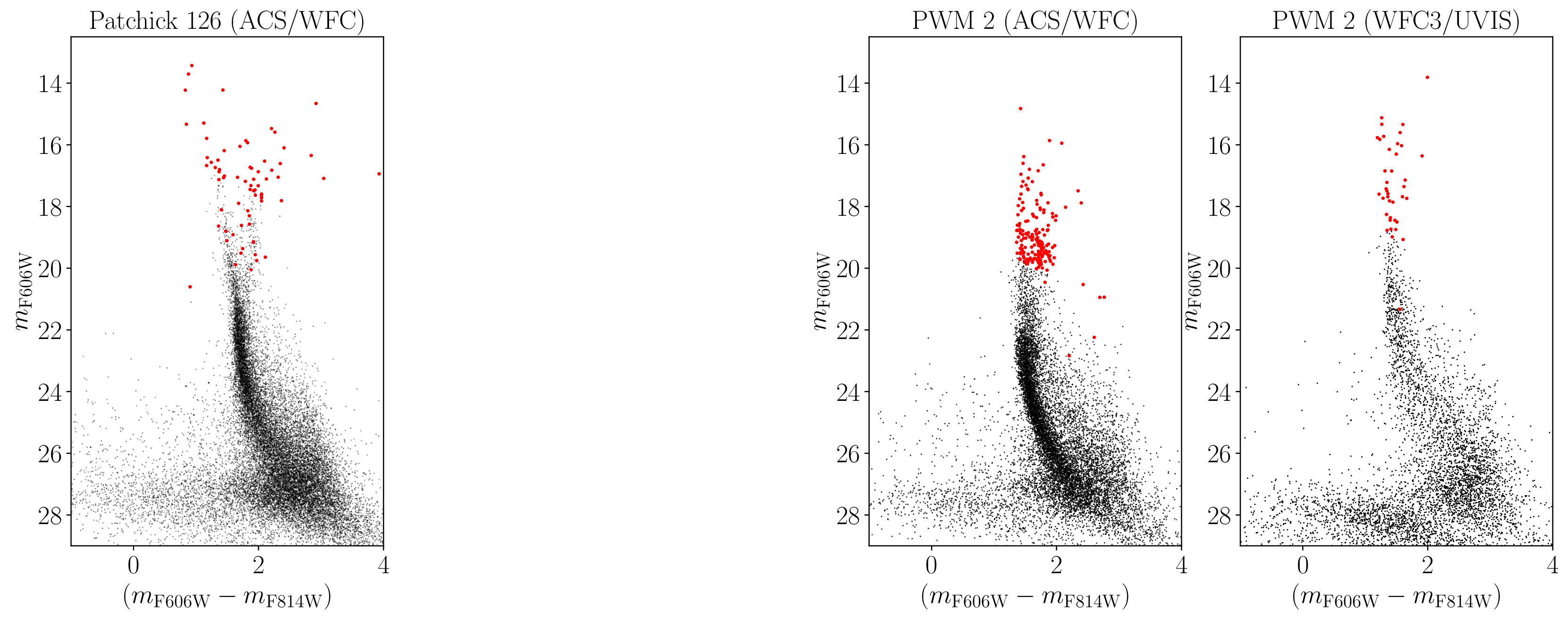}
    \caption{As in Fig.~\ref{fig:cmd0} but for Gran~3, Gran~4, Gran~5, Kim~3, Koposov~1, Koposov~2, Laevens~3, M\~unoz 1, NGC~6749, NGC~7492, Patchick~126, and PWM~2.}
    \label{fig:cmd1}
\end{figure*}

\begin{figure*}
    \centering
    \includegraphics[height=3.6 cm]{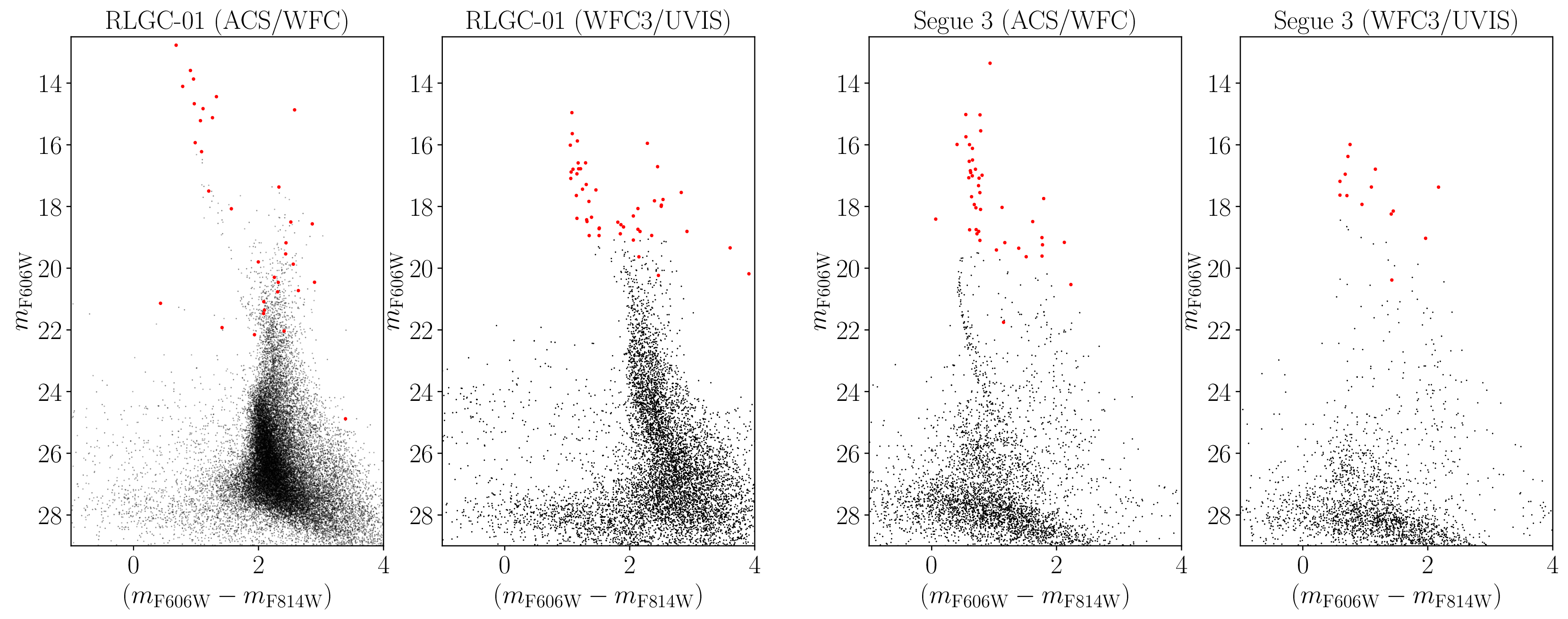}
    \includegraphics[height=3.6 cm]{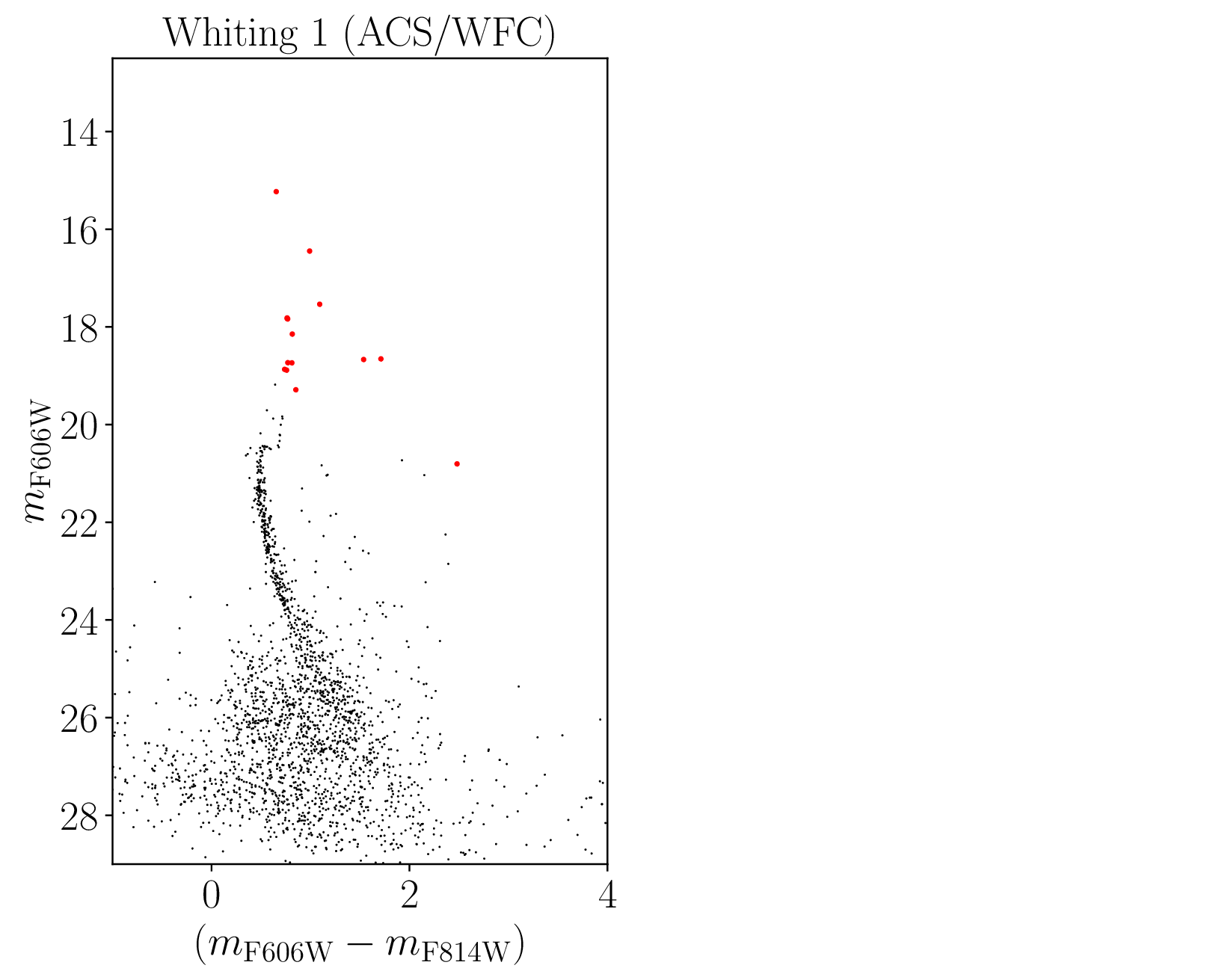}
    \caption{As in Figs.~\ref{fig:cmd0} and \ref{fig:cmd1} but for RLGC-01, Segue~3 and Whiting~1.}
    \label{fig:cmd2}
\end{figure*}

\begin{figure*}
    \centering
    \includegraphics[height=3.6 cm]{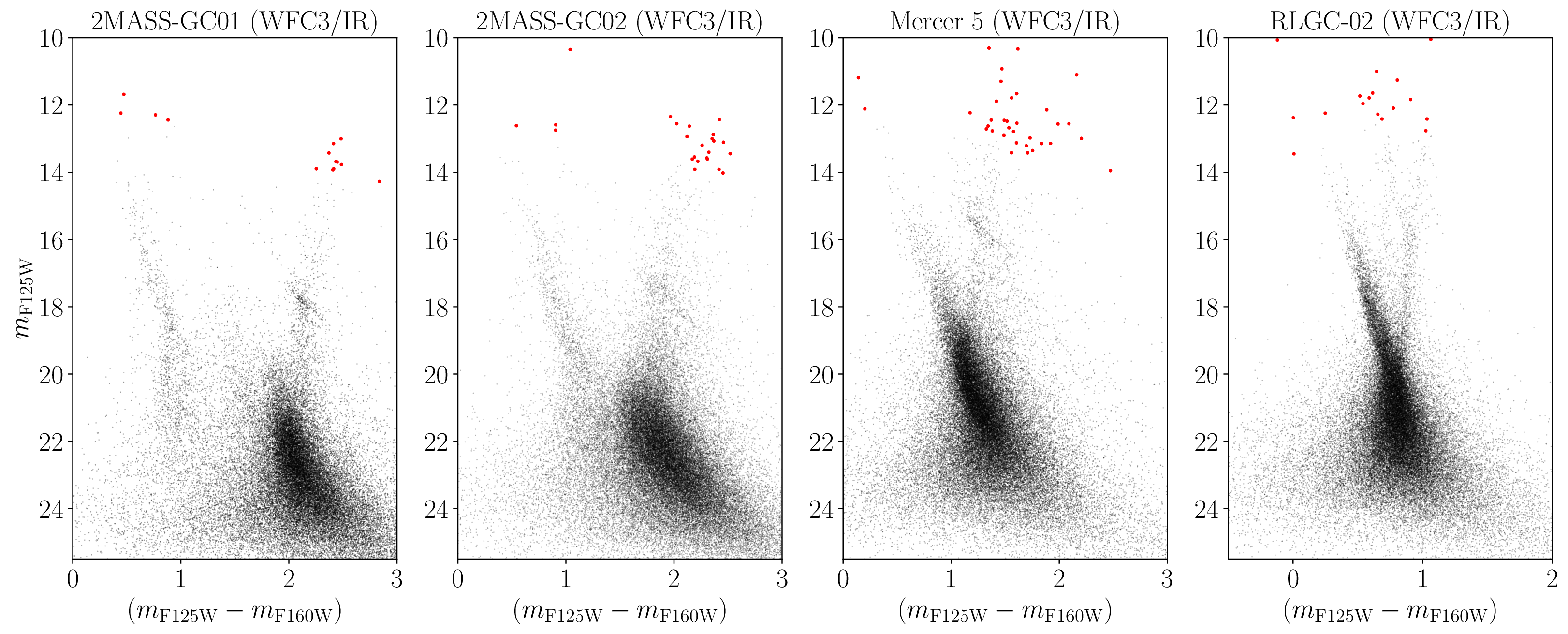}
    \includegraphics[height=3.6 cm]{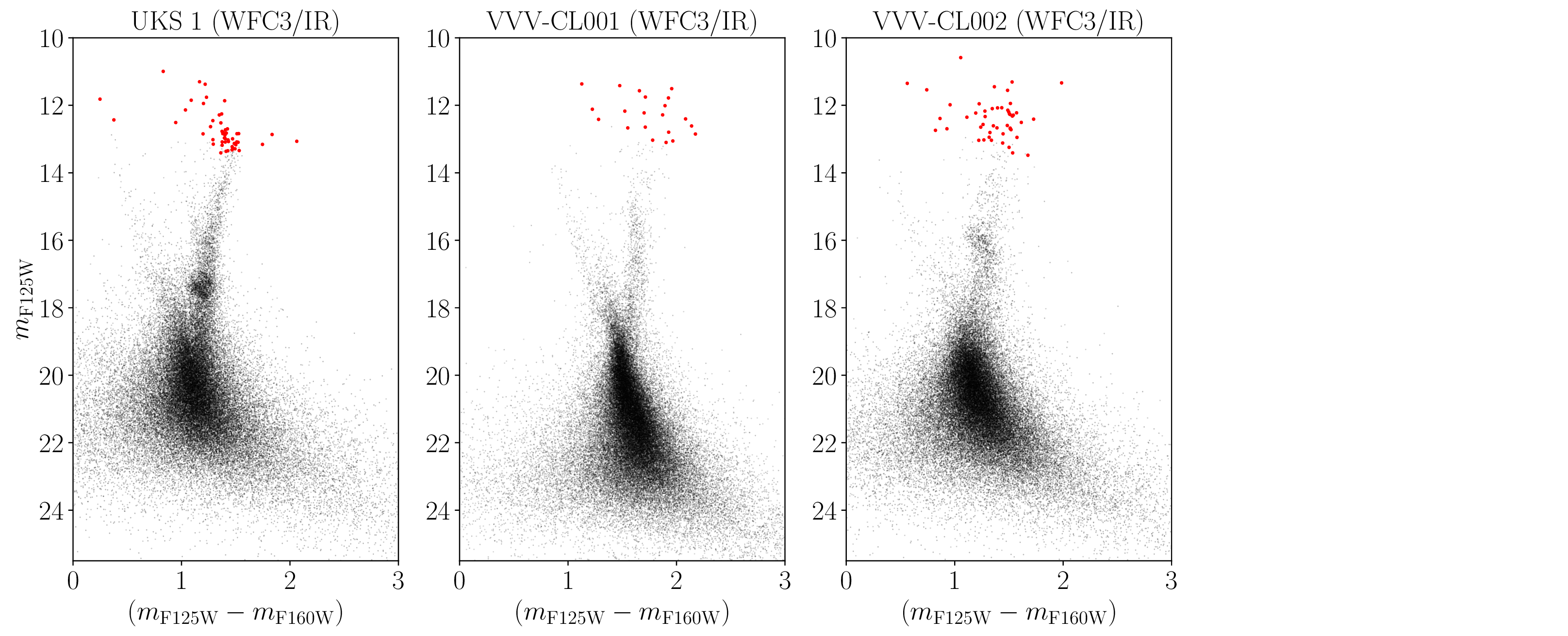}
    \caption{Collection of IR CMDs for 2MASS-GC01, 2MASS-GC02, Mercer~5, RLGC-02, UKS~1, VVV-CL001, and VVV-CL002. Colors have the same meaning as in the other Figures in this Appendix.}
    \label{fig:cmd3}
\end{figure*}

Figures~\ref{fig:cmd0}--\ref{fig:cmd3} present a collection of CMDs for all clusters in our sample.

\clearpage

\end{appendix}

\end{document}